\documentclass[reprint,nofootinbib,superscriptaddress,amsmath,amssymb,aps,prd]{revtex4-2}
\usepackage{amsmath,mathrsfs,amsbsy,color,graphicx,bm,amsthm,amsfonts}
\usepackage{bbm}
\usepackage{times}
\usepackage{units}
\usepackage{dcolumn}
\usepackage{graphicx}
\usepackage{epsfig}
\usepackage{epstopdf}
\usepackage[colorlinks,linkcolor=red,anchorcolor=green,citecolor=blue,CJKbookmarks=True]{hyperref}
\DeclareMathSymbol{\shortminus}{\mathbin}{AMSa}{"39}
\usepackage{mathrsfs}
\usepackage{braket}
\usepackage{amssymb}
\usepackage{txfonts}
\usepackage{float}
\usepackage{enumitem}
\usepackage{multirow}

\begin{document}
\title{\textbf{Observational signature of Lorentz violation in acceleration radiation}}

\author{Yu Tang}
 \affiliation{Department of Physics, Key Laboratory of Low Dimensional Quantum Structures and Quantum \\ Control of Ministry of Education, and Synergetic Innovation Center for Quantum Effects and Applications, Hunan Normal
 University, Changsha, Hunan 410081, P. R. China}

\author{Wentao Liu}
\affiliation{Department of Physics, Key Laboratory of Low Dimensional Quantum Structures and Quantum \\ Control of Ministry of Education, and Synergetic Innovation Center for Quantum Effects and Applications, Hunan Normal
University, Changsha, Hunan 410081, P. R. China}  
        
\author{Jieci Wang}
\email{jcwang@hunnu.edu.cn (Corresponding author)}
\affiliation{Department of Physics, Key Laboratory of Low Dimensional Quantum Structures and Quantum \\ Control of Ministry of Education, and Synergetic Innovation Center for Quantum Effects and Applications, Hunan Normal
University, Changsha, Hunan 410081, P. R. China}

\begin{abstract}
In recent years, Lorentz violation (LV) has emerged as a vibrant area of research in fundamental physics. Despite predictions from quantum gravity theories that Lorentz symmetry may break down at Planck-scale energies, which are currently beyond experimental reach, its low-energy signatures could still be detectable through alternative methods. In this paper, we propose a quantum optical approach to investigate potential LV effects on the acceleration radiation of a freely falling atom within a black hole spacetime coupled to a Lorentz-violating vector field. Our proposed experimental setup employs a Casimir-type apparatus, wherein a two-level atom serves as a dipole detector, enabling its interaction with the field to be modeled using principles from quantum optics. We demonstrate that LV can introduce distinct quantum signatures into the radiation flux, thereby significantly modulating particle emission rates. It is found that while  LV effects are negligible at high mode frequencies, they become increasingly pronounced at lower frequencies. This suggests that detecting LV at low-energy scales may depend on advancements in low-frequency observational techniques or detectors.

\end{abstract}
	\vspace*{0.5cm}
	\maketitle
\section{\textbf{Introduction}}

Recent years have witnessed extensive searches for Lorentz violation (LV) using photons, drawing on observational data from various astrophysical sources. 
These include pulsars \cite{yagi2014strong,shao2014tests,shao2014new}, active galactic nuclei \cite{perennes2020modeling}, gamma-ray bursts \cite{ellis2006robust,zhang2015lorentz,jacob2007neutrinos}, and the cosmic microwave background \cite{leon2017cosmic}. 
By examining low-energy contributions arising from potential LV effects and analyzing high-precision measurements, these studies provide valuable insights into whether Lorentz symmetry is upheld in realistic spacetime scenarios.
However, purely phenomenological constraints derived from astrophysical observations may not be sufficient to reveal the deeper mechanisms through which Lorentz symmetry could be modified—especially in regimes where gravitational and quantum effects converge. 
It is therefore essential to investigate how potential quantum properties of spacetime might alter gravitational theory \cite{kostelecky1989phenomenological,Kostelecky2001,Ashtekar:2021kfp,Lewandowski:2022zce,Zhang:2024khj,Zhang:2024ney,Liu:2024soc,Liu:2024iec,Fu:2024cfk}. 
One promising avenue is to consider that Lorentz violation itself may originate from or be intrinsically linked to such quantum modifications of gravity \cite{choi2015standard,del2023hawking,tian2022probing,agarwalla2023constraining,Filho:2023ycx,Guo:2023nkd,Filho:2024kbq,AraujoFilho:2024rcr,Filho:2024tgy,Hosseinifar:2024wwe,kalb1974classical,yang2023static,duan2023electrically,Liu:2024oas,Liu:2024lve}. 
Along these lines, the bumblebee theory—originally proposed as a vector-tensor extension of Einstein–Maxwell theory—introduces a nonminimally coupled a bumblebee field \cite{kostelecky2004gravity}. 
By acquiring a nonzero vacuum expectation value (VEV), this field spontaneously breaks Lorentz symmetry \cite{casana2018exact,ovgun2019exact,gullu2022schwarzschild,poulis2022exact,maluf2021black,xu2023static,ding2022einstein,Liu:2024axg,Liu:2022dcn,mai2023extended,xu2023bumblebee,zhang2023quasinormal,lin2023quasinormal,chen2023quasinormal,chen2020polarization,wang2022constraint,Liu:2024oeq,Liu:2024wpa,Mai:2024lgk,Chen:2023cjd,AraujoFilho:2025hkm}, offering a theoretical framework in which quantum effects may naturally intertwine with the potential breakdown of Lorentz invariance.

On the other hand, quantum field theory in curved spacetime providing new insights into our understanding of the fundamental laws of the universe \cite{birrell1984,parker2009}.
Two of the most significant predictions and profound insights derived from this theory are the Hawking radiation \cite{hawking1974black,hawking1975particle,hawking1976black,Wu:2022xwy,Wu:2023sye,Liu:2024yrf} and the Unruh effect \cite{unruh1976notes,crispino2008unruh,Wu:2022lmc,Wu:2023spa,Li:2022pwa}.
Both phenomena stem from the same underlying principle as the dynamical Casimir effect described by Moore \cite{moore:1970tmc} that moving boundaries (mirrors) create particles out of vacuum in Minkowski spacetime.
These findings have provided a deeper understanding of the correspondence between moving mirrors and black holes \cite{anderson:2015iga,good:2015jwa}.
Subsequently, Scully { \it et al.} proposed observational setups to detect the thermal radiation associated with event horizons both in flat and curved spacetimes \cite{Scully:2003zz,scully2018quantum}.
These setups are designed to explore  properties of spacetimes by examining the interaction between atomic detectors and quantum fields.
Numerous studies have investigated the dynamical Casimir effect in curved spacetime under various conditions, such as rotation, charge, dark matter, dark energy, braneworld scenarios, and quantum corrections  \cite{Sen:2022cdx,Azizi:2020gff,Sen:2023zfq,Sen:2022tru,Bukhari:2022wyx,Bukhari:2023yuy,Jana:2024fhx,Das:2023rwg,Masood:2024glj}, which provides fresh insights into fundamental physics and driving continued investigations.

In the interaction between a two-level atom and electromagnetic field, a virtual photon is transiently emitted and subsequently reabsorbed, ensuring that the atomic excitation remains a transient event.
However, placing an atom on a freely falling trajectory while positioning a `mirror' at the horizon of a black hole can create relative acceleration between the atom and the field, preventing the atom from recapturing its virtual photon \cite{Scully:2003zz, scully2018quantum}. 
As a result, a genuine photon emerges and is observed by a distant detector. 
Given the observable effects of this accelerated radiation, a natural question arises: can we utilize this phenomenon to search for potential Lorentz violation in spacetime?
In this work, we introduce a novel approach to investigate LV by examining the accelerated radiation emission of freely falling atoms in a black hole with Lorentz-violating vector field. It is shown that the coupling between curved spacetime and the Lorentz-violating vector field significantly affects particle emission. The LV effect may manifest through the altered radiation intensity of atoms, either enhancing or reducing it.  In addition, the LV effect is negligible at high frequencies but becomes more pronounced at lower frequencies. 

The organization of the paper is as follows.
In Sec. \ref{sec2}, We discuss the essential background for understanding our working principle, focusing on the Einstein-Bumblebee gravity model and the near-horizon physics of black holes, which drive the acceleration radiation.
In Sec. \ref{sec3}, we calculate the particle spectrum of accelerated radiation in the near-horizon region and analyze the results.
The conclusions and outlooks are presented in Sec. \ref{sec4}.
Furthermore, we adopt the metric signature $(-~+~+~+)$, with all quantities expressed in natural units ($\hbar=c=1$).

\section{BLACK HOLE SPACETIME AND NEAR-HORIZON CONFORMAL QUANTUM MECHANICS EQUATION}\label{sec2}
	\subsection{ Lorentz-violating black hole spacetime}
		The action for the bumblebee field $B_\mu$ coupled to gravity can be described as \cite{casana2018exact}
		\begin{equation}
		\begin{aligned}\label{Action}
		\mathcal{S}_B=&\int d^4x \sqrt{-g}\left[\frac{1}{2\kappaup}\left(R-2\Lambda\right)+\frac{\varrho}{2\kappa} B^\mu B^\nu R_{\mu\nu} \right.\\
		&~~~\left. -\frac{1}{4}B^{\mu\nu}B_{\mu\nu}-V\left(B^\mu B_\mu\pm b^2\right)\right],
		\end{aligned}
		\end{equation}
			where $ \kappaup=8\pi G_N $ is the gravitational coupling constant which can be set to $ G_N =1$ without loss of generality.
			$ \Lambda $ is the cosmological constant and $ \varrho $ is the real coupling constant which controls the non-minimal gravity interaction to the bumblebee field $B_\mu$.
			It is worth noting that the potential $V$, chosen to ensure a non-zero VEV for the bumblebee field $\langle B_\mu\rangle\equiv b_\mu$, triggers spontaneous Lorentz symmetry breaking, reaching its minimum at $B_\mu B^\mu= \mp b^2$, where $b$ is a positive real constant,  and the $ \pm $ sign implies that $ B_\mu $ is timelike or spacelike, respectively.

		Toking of the variational $g_{\mu\nu}$ and $B_\mu$ yields the gravitational equation and bumblebee field equation:
		\begin{align}
		& R_{\mu\nu}-\frac{1}{2}g_{\mu\nu}\left(R-2\Lambda\right)=\kappa T^{B}_{\mu\nu} \label{EinsteinEQ}, \\ \label{VEq}
		& \nabla^\mu B_{\mu\nu}=2V'B_\nu-\frac{\varrho}{\kappa}B^\mu R_{\mu\nu}.
		\end{align}
		$T^{B}_{\mu\nu}$ is the bumblebee energy momentum tensor, which have the following form:
		\begin{align}
		T_{\mu\nu}^{B}=&B_{\mu\alpha}B^{\alpha}{}_{\nu}-\frac{1}{4}g_{\mu\nu}B^{\alpha\beta}B_{\alpha\beta}-g_{\mu\nu}V+2B_{\mu}B_{\nu}V^{\prime}\notag
		\\&+\frac{\varrho}{\kappa}\left[\frac{1}{2}g_{\mu\nu}B^{\alpha}B^{\beta}R_{\alpha\beta}-B_{\mu}B^{\alpha}R_{\alpha\nu}-B_{\nu}B^{\alpha}R_{\alpha\mu}\right]\notag
		\\&+\frac{1}{2}\nabla_{\alpha}\nabla_{\mu}\left(B^{\alpha}B_{\nu}\right)+\frac{1}{2}\nabla_{\alpha}\nabla_{\nu}\left(B^{\alpha}B_{\mu}\right)\notag
		\\&-\frac{1}{2}\nabla^{2}\left(B_{\mu}B_{\nu}\right)-\frac{1}{2}g_{\mu\nu}\nabla_{\alpha}\nabla_{\beta}\left(B^{\alpha}B^{\beta}\right)\label{TBab}.
		\end{align}

		An exact spherically symmetric black hole solution has been constructed by Casana {\it et al.} \cite{casana2018exact}, and it is referred to as  Schwarzschild-like black holes.
		The bumblebee field $ B_\mu $ can be written as
		\begin{align}
		B_\mu=b_\mu=\left\{0,b\sqrt{\frac{(1+\ell)}{F(r)}},0,0\right\},
		\end{align}
		and the line element of spacetime is given by
		\begin{align}
		            &ds^{2}=-F(r)dt^{2}+\frac{1+\ell}{F(r)}dr^{2}+r^{2}d\theta^{2}+r^{2}\sin^{2}\theta d\phi^{2},\label{eq:metric}
		\end{align}
		where metric function is $ F(r)=1-2M/r $ and the symbol $ \ell=\varrho b^2 $ represents the Lorentz violation parameter.
		From a gravitational perspective, such solution are supported by an anisotropic energy-momentum tensor, considered as the manifestation of the bumblebee field within spacetime geometry.

	\subsection{Geodesic equations}
\begin{figure}[h]
		\centering 
		\includegraphics[width=0.8\linewidth]{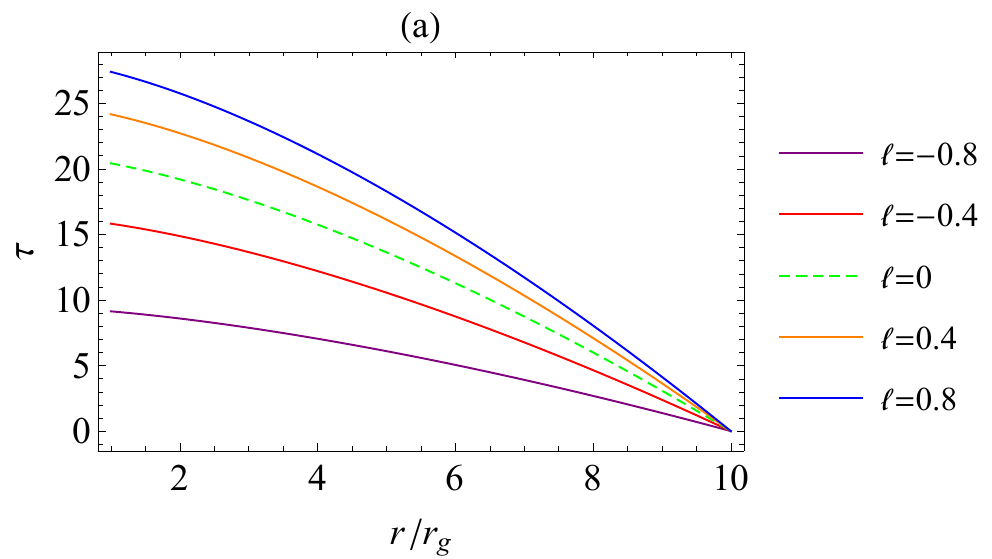} 
		\includegraphics[width=0.8\linewidth]{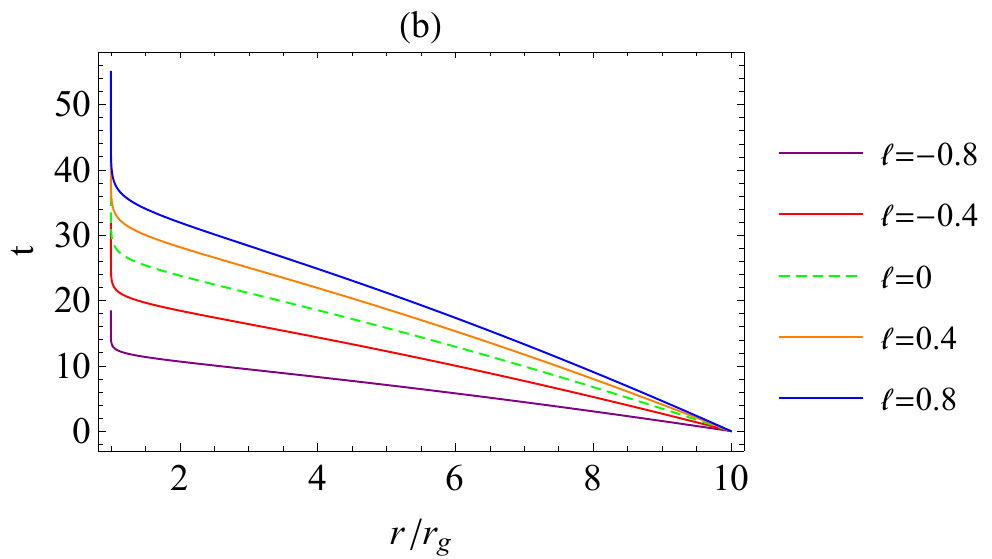}   
		\caption{Impact of $\ell$ on the behavior of proper time $\tau$ and coordinate time $t$ against the normalization radial coordinate $r/r_g$.
		We assume the atom begins to fall freely to the black hole from the asymptotically flat spacetime.}
		\label{figt}
		\end{figure}

		The spacetime trajectories for free-fall motion of the atoms in a background metric $g_{\mu\nu}$ are described by the geodesic equations \cite{chandrasekhar1998mathematical}
		\begin{equation}
		    \frac{d^2x^\mu}{d\tau^2}+\Gamma_{\rho\sigma}^\mu\frac{dx^\rho}{d\tau}\frac{dx^\sigma}{d\tau}=0, \label{eq:geo}
		\end{equation}
		where $\Gamma_{\rho\sigma}^\mu$ are the Christoffel connections.
		For a static and spherically symmetric metric defined by Eq.~(\ref{eq:metric}), we can restricting the motion of atom to an equatorial plane and take $\theta=\pi/2$, giving $\dot{\theta}=\dot{\phi}=0$.
		This procedure reduces the problem to finding the geodesics with initial conditions defined by
		\begin{equation}
		    e=F(r)\frac{dt}{d\tau}, \label{eq:eng}
		\end{equation}
		where $e$ represents the energy per unit mass.
		For a free fall atom from a fiducial point, the conserved quantity $e$ give the initial conditions.
		Hence one obtains the following conservation equations
		\begin{equation}
		    \left(\frac{dr}{d\tau}\right)^2=\frac{e^2-F(r)}{1+\ell},\quad\left(\frac{dr}{dt}\right)^2=\frac{\left(\frac{F(r)}{e}\right)^2[{e}^2-F(r)]}{1+\ell}. \label{eq:con}
		\end{equation}
		We can integrate these equations to get the atom’s proper time $\tau$ and the coordinate time $t$ in terms of the radial variable $r$
		\begin{equation}
		    \tau=-\int_{r_i}^{r_f}dr\frac{\sqrt{1+\ell}}{\sqrt{e^2-F(r)}},t=-\int_{r_i}^{r_f}dr\frac{e\sqrt{1+\ell}}{F(r)\sqrt{e^2-F(r)}}, \label{eq:time}
		\end{equation}
		where $r_{i}$ and $r_{f}$ are the the initial and final positions of the atom, respectively.
		In particular, if the atom begins to fall freely to the black hole from the asymptotically flat spacetime, one can take $e=1$, and it is possible to analytically solve the equation for $\tau(r)$ and $t(r)$.
		These quantities are plotted in Figs. \ref{figt}(a)-\ref{figt}(b).
		
		The behaviors of $t(r)$ and $\tau(r)$ demonstrate the typical Schwarzschild-type character.
		It can be observed that the Lorentz violation influences the position of the black hole's event horizon.
		
	\subsection{Near-horizon conformal quantum mechanics equation} \label{cqm}
		Then we compute the solution of the Klein-Gordon equation in the black hole background.
		In the subsequent calculation, it is assumed that the angular dependence of the modes can be neglected.
		The near-horizon analysis will focus on the functional dependence of the fields in the neighborhood of the outer event horizon at $r=r_{g}=2M$ \cite{camblong2005black,camblong2005semiclassical}.
		The wave equation for a massless Klein-Gordon field in the minimal coupling  is given by \cite{birrell1984}
		\begin{align}
		            &\frac{1}{\sqrt{-g}}\partial_{\mu}\left(\sqrt{-g}g^{\mu\nu}\partial_{\nu}\Phi\right)=0. \label{eq:kg}
		        \end{align}
		With the choice of metric given in Eq.~(\ref{eq:metric}), we  consider the following mode expansion of the scalar field \cite{camblong2020near}
		\begin{equation}
		    \Phi(t,r)=\sum\left[\hat{a}\phi(r,t)+H.c.\right], \label{eq:exp}
		\end{equation}
		where $\hat{a}$ is the field annihilation operator, H.c. means hermitian conjugate, and $\phi$ constitute a complete set of orthonormal solutions to Eq.~(\ref{eq:kg}).
		The use of black hole coordinates selects these special modes for expansion.
		This includes the existence of an associated Boulware vacuum $\left|0_{B}\right\rangle$ \cite{boulware1975quantum} such that
		\begin{equation}
		    \hat{a}|0_B\rangle=0, \label{eq:bwc}
		\end{equation}
		which asymptotically manifested as a Minkowski vacuum at infinity.
		
		The Eq.~(\ref{eq:kg}) has the following separable variable solutions in the coordinates
		\begin{equation}
		    \phi(t,r)=\chi(r)u(r)e^{-i\nu t}, \label{eq:svs}
		\end{equation}
		where the time dependence involves frequencies $\nu$.
		In particular, the choice of $\chi(r)=[F(r)]^{-1/2}$ reduces the radial part of the Klein-Gordon equation to its standard form
		\begin{equation}
		    u^{\prime\prime}(r)+V_{(D)}(r;\nu)u(r)=0, \label{eq:skg}
		\end{equation}
		where $V_{(D)}$ is effective potential, and its full form is given in Appendix \ref{appa}.
		We can examine the behavior of the modes $u(r)$ near the horizen $\mathcal{H}$, $r\sim r_g$.
		Utilizing the shifted variable $x=r-r_{g}$, the expansions of $F(r)$ and its derivatives are given by
		\begin{align}
		            &F(r)\overset{(\mathcal{H})}{\operatorname*{\sim}}F_g^{\prime}x\left[1+\mathcal{O}(x)\right],\notag   \\
		            &F^{\prime}(r)\overset{(\mathcal{H})}{\operatorname*{\sim}}F_g^{\prime}[1+\mathcal{O}(x)],\notag   \\
		            &F^{\prime\prime}(r)\overset{(\mathcal{H})}{\operatorname*{\sim}}F_g^{\prime\prime}\left[1+\mathcal{O}(x)\right], \label{eq:taly}
		\end{align}
		where $F_g^{\prime\prime}\equiv F^{\prime\prime}(r_g)$ and the notation $\overset{(\mathcal{H})}{\operatorname*{\operatorname*{\sim}}}$ means the hierarchical expansion near the horizon.
		Using the near-horizon expansion (\ref{eq:taly}), we can simplify the effective potential in Eq.~(\ref{eq:skg}) and obtain (please see Appendix \ref{appa} for detail)
		\begin{equation}
		    u^{\prime\prime}(x)+\frac{\lambda_{\mathrm{eff}}}{x^2}\left[1+\mathcal{O}(x)\right]u(x)=0. \label{eq:cqm}
		\end{equation}
		Eq.~(\ref{eq:cqm}) means that the dominant physics near the horizon is driven by the interaction
		\begin{equation}
		    V_{\mathrm(D)}(x)=\frac{\lambda_{\mathrm{eff}}}{x^2},\quad\lambda_{\mathrm{eff}}=\frac{1}{4}+\Theta^2,\quad\Theta=\frac{\nu\sqrt{1+\ell}}{F_g^{\prime}}.
		\end{equation}
		This is the well-known long-range representative of conformal quantum mechanics (CQM) \cite{camblong2003anomaly}.

\section{PARTICLE SPECTRUM OF ACCELERATION RADIATION}\label{sec3}
		For the sake of clarity, we draw a schematic picture.
		\begin{figure}[h]
		    \centering
		    \includegraphics[width=0.33\textwidth]{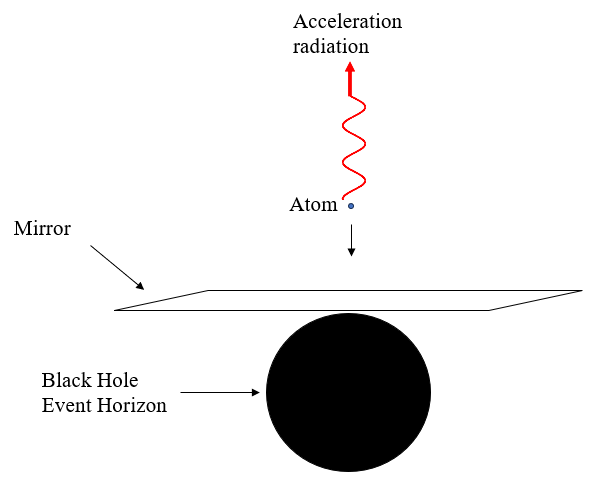}
		    \caption{Schematic view of acceleration radiation emitted from a two-level atom falling into a black hole.
		    A “mirror” is held at the event horizon which shields infalling atoms from the Hawking radiation.
		    The relative acceleration between the atoms and the field yields generation of acceleration radiation.}
		    \label{fig:sch}
		\end{figure}
		As shown in Fig. \ref{fig:sch}, a BH at the center  and a mirror held at the event horizon, which shields infalling atoms from the Hawking radiation.
		A two-level dipole atom, which acts as the detector, falls freely towards the black hole described by the metric Eq.~(\ref{eq:metric}).
		The relative acceleration between the atoms and the field modes leads to the generation of acceleration radiation which is received by an observer at asymptotic infinity \cite{scully2018quantum}.
		Since there is no Hawking flux received by the asymptotic observer, we first assume the field to be in Boulware vacuum state Eq.~(\ref{eq:bwc}).
		As the atom falls towards the black hole, it detects radiation by going to the excited state and emitting a photon.
		This simultaneous atomic excitation and photon emission originates from acceleration and is determined by the Unruh effect \cite{unruh1976notes,crispino2008unruh}.
		Using time-dependent perturbation theory, the excitation probability for the atom to make a transition from a ground state $\left|b\right\rangle$ to excited state $\left|a\right\rangle$ while emitting a photon of frequency $\nu$ is given by
		\begin{equation}
		    P_{exc}=\frac{1}{\hbar^{2}}\left|\int d\tau\langle1_{\nu},a|V_{I}(\tau)|0,b\rangle\right|^{2}, \label{eq:p}
		\end{equation}
		where $\tau$ is the atom’s proper time and $\left|1_{\nu}\right\rangle$ represents the one-photon mode with frequency $\nu$.
		
		The relevant interaction potential $V_{I}(\tau)$ is given by \cite{scully2018quantum}
		\begin{equation}
		    V_I(\tau)=\hbar g\left[\hat{a}_{\nu}\phi(r(\tau),t(\tau))+H.c.\right]\left(\sigma_-e^{-i\omega\tau}+H.c.\right), \label{eq:v}
		\end{equation}
		where $\sigma_{-}$ is atomic lowering operator, $\omega$ is the atom frequency and $g$ denotes the strength of the interaction.
		In addition, $\phi(r,t)$ can be obtained from the Klein-Gordon equation as in subsection \ref{cqm}, and $\hat{a}_{\nu}$ stands for the associated annihilation operators.
		
		It is clear from the expression of $P_{exc}$ in Eq.~(\ref{eq:p}) that only the term corresponding to $\hat{a}_{\nu}^\dagger\sigma_-^\dagger$ gives non-zero contribution to the probability.
		This enables us write down Eq.~(\ref{eq:p}) as
		\begin{equation}
		    P_{exc}=g^{2}\left|\int d\tau\phi^{*}(r(\tau),t(\tau))e^{i\omega\tau}\right|^{2}. \label{eq:p1}
		\end{equation}
		
		The field modes required to calculate the excitation probability $P_{exc}$ in in Eq.~(\ref{eq:p1}) can be obtained using the CQM Eq.~(\ref{eq:cqm}).
		We will select the atomic outgoing radiation wave from the neighborhood of the event horizon \cite{camblong2020near}, which is given by
		\begin{equation}
		    u(x)=x^{\frac{1}{2}+\sqrt{\frac{1}{4}-\lambda}}=\sqrt{x}x^{i\Theta}, \label{eq:ux}
		\end{equation}
		where $\Theta=\nu\sqrt{1+\ell}/F_g^{\prime}$.
		Because the angular dependence is not needed, we combine all the factors together to get the field mode $\phi(r,t)=\chi(r)u(r)e^{-i\nu t}$.
		Then, in the near-horizon expansion one obtains
		\begin{equation}
		    \chi(r)=[F(r)]^{-1/2}\overset{(\mathcal{H})}{\operatorname*{\sim}}\frac{1}{\sqrt{x}\sqrt{F_{g}^{\prime}}}(1+\mathcal{O}(x)). \label{eq:xr}
		\end{equation}
		Thus, we can find that
		\begin{equation}
		    \phi(r,t)\overset{(\mathcal{H})}{\operatorname*{\sim}}\frac{1}{\sqrt{F_{g}^{\prime}}}{x}^{i\Theta}e^{-i\nu t}\longrightarrow\phi(r,t)\overset{(\mathcal{H})}{\operatorname*{\sim}}{x}^{i\Theta}e^{-i\nu t}=e^{-i\nu(t-\sqrt{1+\ell}\ln x/F_{g}^{\prime})}. \label{eq:bohs}
		\end{equation}
		In the last step of Eq.~(\ref{eq:bohs}), by removing the additional constant factor in the leading near-horizon approximation, the pure-phase outgoing wave near the event horizon is extracted \cite{camblong2020near}.
		
		The spacetime trajectories for free-fall motion of the atom in a background metric $g_{\mu\nu}$ are described by Eq.~(\ref{eq:time}). Therefore, the integration of $\tau$ and $t$ can now be performed by using a Taylor expansion around the event horizon as a function of the near-horizon variable $x=r-r_g$.
		Up to first order in $x$, one obtains
		\begin{align}
		            & \hspace{1cm}\tau=-\frac{\sqrt{1+\ell}}{e}x+\mathrm{const.}+\mathcal{O}(x^2),\label{eq:taylp}\\
		            &t=-\frac{\sqrt{1+\ell}}{F_g^{\prime}}\ln x-C\sqrt{1+\ell}x+\mathrm{const.}+\mathcal{O}(x^2), \label{eq:taylt}
		\end{align}
		where $C$ is a constant dependent on the conserved quantities:
		\begin{equation}
		    C=\frac{1}{2}\left[\frac{1}{e^2}-\frac{F_g^{\prime\prime}}{(F_g^{\prime})^2}\right]. \label{eq:ccc}
		\end{equation}
		In particular, we can neglect the $\mathcal{O}(x^2)$ terms near the horizon as their contribution becomes insignificant.
		
\begin{figure}[h]
		\centering 
		\includegraphics[width=0.48\linewidth]{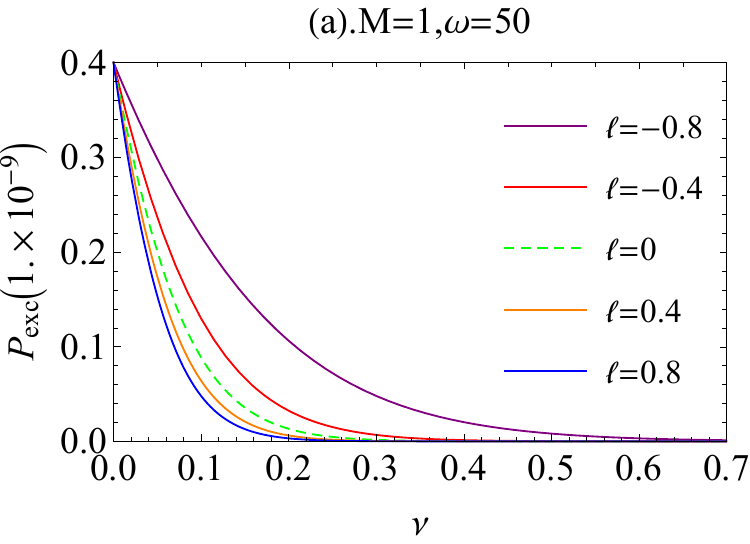}    
		\includegraphics[width=0.48\linewidth]{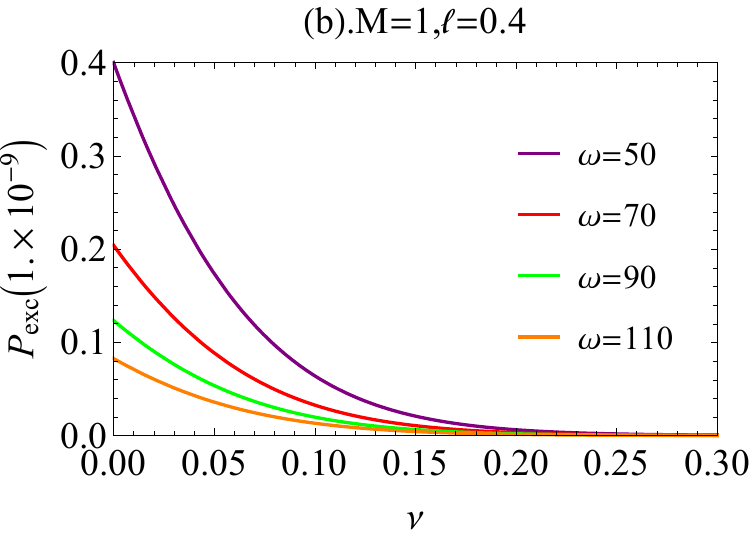}    
		\includegraphics[width=0.48\linewidth]{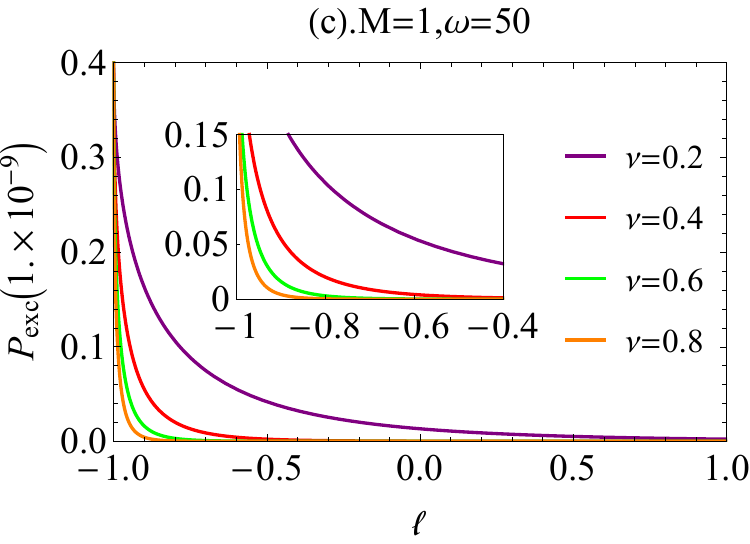}    
		\includegraphics[width=0.49\linewidth]{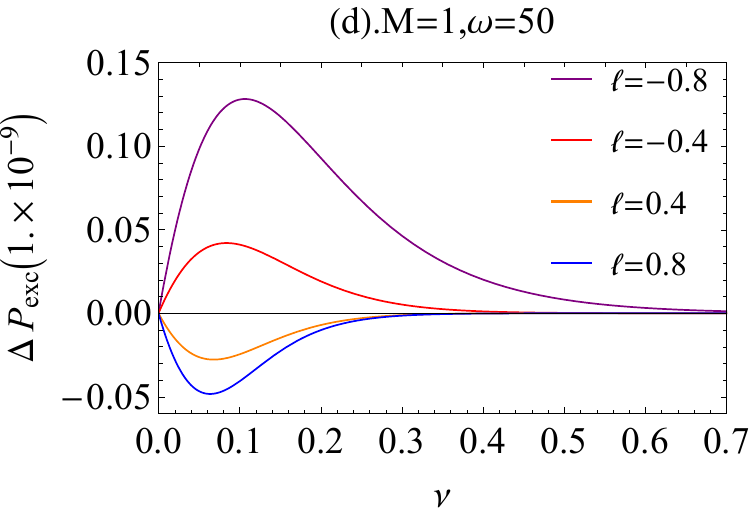}   
		\caption{(a) The radiation intensity $P_{exc}$ as a function of mode frequencies $\nu$ for different  Lorentz violation parameter $\ell$; (b) $P_{exc}$ as a function of mode frequencies $\nu$ for different  atomic frequency $\omega$;	(c) Radiation intensity $P_{exc}$ as a function of the Lorentz violation parameter $\ell$;
		(d) $\Delta P_{exc}=P_{exc}(\ell)-P_{exc}(0)$ as a function of mode frequencies $\nu$. Here we take $g=10^{-3}$ for clarity.}
		\label{fig3}
		\end{figure}

\begin{figure}[h]
		\centering 
		\includegraphics[width=0.75\linewidth]{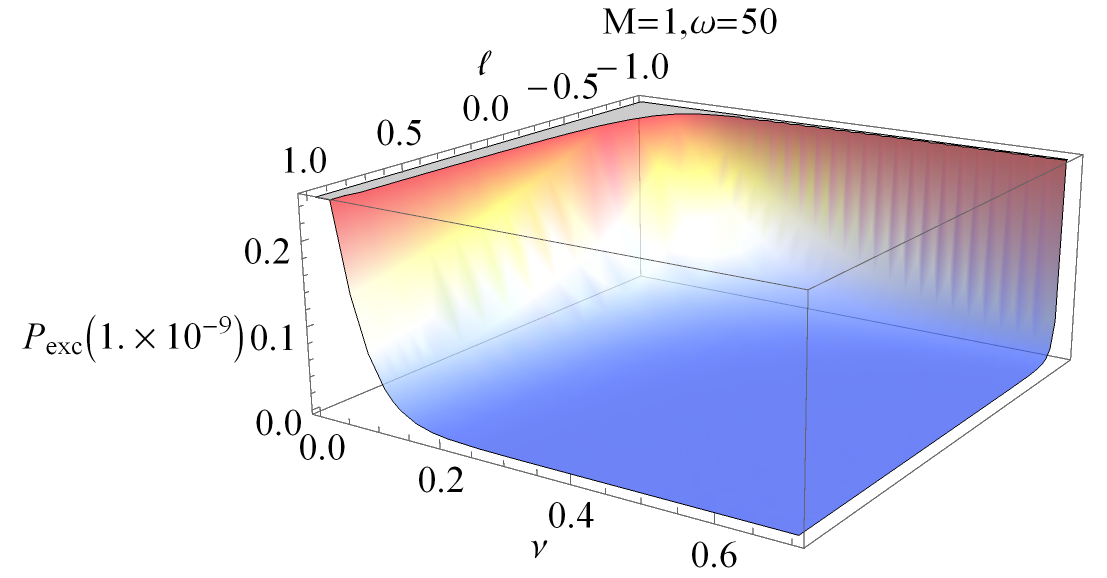} 
		\caption{Radiation intensity $P_{exc}$ as a function of  mode frequencies $\nu$ and the Lorentz violation parameter $\ell$.}
		\label{fig4}
		\end{figure}

		Using the above results, one can calculate the excitation probability $P_{exc}$.
		Substituting Eqs.~(\ref{eq:bohs}), (\ref{eq:taylp}) and (\ref{eq:taylt}) into Eq.~(\ref{eq:p1}), the equation yields
		\begin{align}
		    P_{exc}&=\frac{g^2 \left(1+\ell\right)}{e^2}\left|\int_0^{x_f}dxx^{-i\Theta}e^{i\nu\sqrt{1+\ell}(-\ln x/F_g^{\prime}-Cx)}e^{-i\omega\sqrt{1+\ell} x/e}\right|^2 \notag \\
		    &=\frac{g^2 \left(1+\ell\right)}{e^2}\left|\int_0^{x_f}dxx^{-i\sigma}e^{-isx}\right|^2, \label{eq:p2}
		\end{align}
		where
		\begin{align}
		& \hspace{1.5cm}\sigma=2\Theta=\frac{2\nu\sqrt{1+\ell}}{F_g^{\prime}}=2r_g\nu\sqrt{1+\ell}\label{eq:sigma}, \\
		&s=\sqrt{1+\ell}\left(C\nu+\frac{\omega}{e}\right)=\frac{\nu\sqrt{1+\ell}}{2}\left[\frac{1}{e^2}-\frac{F_g^{\prime\prime}}{(F_g^{\prime})^2}\right]+\frac{\omega\sqrt{1+\ell}}{e}. \label{eq:sss}
		\end{align}
		In Eq.~(\ref{eq:p2}), $x_f$ represents the boundary of a region where the near-horizon approximation is valid.
		In the limit when $s\gg\sigma$, we can push the upper limit of the integration to infinity \cite{camblong2020near} and evaluate the integral Eq.~(\ref{eq:p2}) using 
		\begin{equation}
		    \int_0^\infty dxx^{2i\nu}e^{ix}=-\frac{\pi e^{-\pi\nu}}{\sinh{(2\pi\nu)}\Gamma{(-2i\nu)}},
		\end{equation}
		where $\Gamma(z)$ is the gamma function. Then we find
		\begin{align}
		    P_{exc}&\approx\frac{2\pi g^2\sigma\left(1+\ell\right)}{e^2s^2}\frac{1}{e^{2\pi\sigma}-1}\notag \\
		    &=\frac{4\pi g^2 r_g\nu\sqrt{1+\ell}}{\omega^2\left(1+Ce\nu/\omega\right)^2}\frac{1}{e^{2\pi\sigma}-1}\notag \\
		    &\approx\frac{4\pi g^2r_g\nu\sqrt{1+\ell}}{\omega^2}\frac{1}{e^{4\pi r_g\nu\sqrt{1+\ell}}-1}, \label{eq:pf}
		\end{align}
		where the approximation $\omega\gg\nu$ is enforced in the last step.
		The frequency hierarchy $\omega\gg\nu$ is a “geometrical optics” approximation for the fall of the atom \cite{scully2018quantum}.
		We are now able to examine the effects of Lorentz violation and other parameters on acceleration radiation. The results are illustrated in Figs. \ref{fig3} and \ref{fig4}.
		
Considering the aforementioned formulation, it is intriguing to observe the emission of radiation in a Boulware vacuum state, which conventionally exhibits no Hawking flux at asymptotic infinity. Furthermore, it is found that the frequency hierarchy $\omega\gg\nu$ means a semiclassical treatment of the particle geodesics as well-defined classical paths.
		At the same time, in the process of calculating the excitation probability $P_{exc}$, we use the reduction of the denominator $(1+Ce \nu/\omega)^2\approx1$. This means the final result for $P_{exc}$ is independent of the numerical factor $C$.
		In addition, the derivation indicates that the final expression for $P_{exc}$ completely conforms to the Planck distribution at the Hawking temperature. In other words, it exhibits behavior similar to that of Hawking radiation itself.
		
		Recent studies \cite{azizi2021quantum,chakraborty2019detector} have showed thermal production of particles depicting either a Planckian or a Bose-Einstein (BE) distributions in pure black holes. Here we illustrate in Fig. \ref{fig3}(a) a thermal Bose-Einstein-type distribution for field quanta, indicating that Lorentz violation maintains thermality.  We further sought to quantify the impact of the atomic transition frequency  $\omega$ on the excitation probability, as illustrated in Fig. \ref{fig3}(b). According to the Unruh effect\cite{unruh1976notes,crispino2008unruh}, an increase in  frequency $\omega$ tends to suppress atomic excitation, consequently diminishing emission efficiency. This observation suggests that Lorentz violation maintains the key characteristics of the Unruh effect.  As shown in Figs. \ref{fig3}(c) and \ref{fig3}(d), the effects of Lorentz violation are negligible at high mode frequencies but become significant at low frequencies. We contend that this is due to the low energy of these fields enhances the significance of the Lorentz-violating field’s non-zero vacuum expectation value. Finally, as shown in Fig. \ref{fig4}, the smaller the value of $\ell$ or $\nu$, the greater the radiation intensity $P_{exc}$ is.

\section{DISCUSSIONS AND CONCLUSIONS}\label{sec4}
In this paper, we examine the emission of acceleration radiation by a freely-falling atom in a Lorentz-violating black hole. Within the near-horizon approximation, this radiation displays characteristics analogous to those of Hawking radiation. We find that the coupling between curved spacetime and the Lorentz-violating vector field can significantly influence particle emission. Specifically, at lower field energies, the non-zero vacuum expectation value of the Lorentz-violating field becomes more pronounced, leading to a more evident manifestation of Lorentz violation at lower field mode frequencies. This implies that the detection of Lorentz violation at low-energy scales may be contingent upon advancements in low-frequency observational techniques or detectors.

				The analysis method discussed in this paper is also applicable to models in which a black hole is coupled to a Lorentz-violating tensor field \cite{kalb1974classical,yang2023static,duan2023electrically,Liu:2024oas,Liu:2024lve}.
		In such cases, the spacetime structure will deviate from that described by the Eq.~(\ref{eq:metric}).
		Since Lorentz violation influences the position of the black hole's event horizon, leading to a deviation from the Schwarzschild radius.
		Future research can gain a deeper insight into the quantum properties of spacetime by comparing the phenomena of Lorentz violation in different fields within Lorentz-violating spacetimes, induced by the coupling of vectors and tensors.
		In addition, exploring the connections between black hole topology classification \cite{Wei:2022dzw,Wei:2022dzw,Wu:2022whe,Wu:2023xpq,Wu:2023fcw,Wu:2023meo,Wu:2024rmv,Wu:2024asq} and Lorentz symmetry breaking represents a promising direction for further investigation. For instance, do black hole spacetimes with identical topological numbers exhibit consistent observable effects? This could further illuminate the profound connections between  Lorentz-violating field and quantum gravity theories.

\acknowledgments
This work was supported by the National Natural Science Foundation of China under Grants  No. 12475051,  No. 12375051, and No. 12421005; the science and technology innovation Program of Hunan Province under grant No. 2024RC1050; the Natural Science Foundation of Hunan Province under grant No. 2023JJ30384;  the innovative research group of Hunan Province under Grant No. 2024JJ1006; and the Postgraduate Scientific Research Innovation Project of Hunan Province  under grant No. CX20240531.

\appendix
\section{Derivation of CQM equation} \label{appa}
		For the class of metrics Eq.~(\ref{eq:metric}), the wave equation for a massless Klein-Gordon field in the minimal coupling (\ref{eq:kg}) reduces to
		\begin{equation}
		    -\frac{1}{F(r)}\frac{\partial^{2}\Phi}{\partial t^{2}}+\frac{F(r)}{1+\ell}\frac{\partial^{2}\Phi}{\partial r^{2}}+\frac{F^{\prime}(r)}{1+\ell}\frac{\partial\Phi}{\partial r}=0,\label{eq:app1}
		\end{equation}
where it is assumed that the angular dependence of the modes can be neglected. Considering the mode expansion of the following scalar field, one obtains
		\begin{equation}
		    \Phi(t,r)=\sum\left[\hat{a}\phi(r,t)+H.c.\right], \label{eq:app2}
		\end{equation}
		with the following ansatz 
		\begin{equation}
		    \phi(t,r)=\chi(r)u(r)e^{-i\nu t}. \label{eq:app3}
		\end{equation}
		In particular, the choice of the radial function
		\begin{equation}
		    \chi(r)=\exp\left(-\frac{1}{2}\int\frac{F(r)^{\prime}}{F(r)}dr\right)=[F(r)]^{-1/2}\label{eq:app4}
		\end{equation}
		reduces the radial part of Eq.~(\ref{eq:app1}) to the canonical form
		\begin{equation}
		    u^{\prime\prime}(r)+V_{(D)}(r;\nu)u(r)=0, \label{eq:app5}
		\end{equation}
		where
		\begin{equation}
		    V_{(D)}(r;\nu)=\frac{1}{F^{2}}\left[\nu^{2}(1+\ell)+\frac{F^{\prime2}}{4}\right]-\frac{F^{\prime\prime}}{2F}.\label{eq:app6}
		\end{equation}
		Using the near-horizon expansion (\ref{eq:taly}), the leading orders of each one of the terms on the right-hand side of Eq.~(\ref{eq:app6}) become
		\begin{align}
		    V_{(D)}(r;\nu)&\overset{(\mathcal{H})}{\operatorname*{\sim}}\frac{1}{(F_{g}^{\prime})^{2}}\left[\nu^{2}(1+\ell)+\frac{(F_{g}^{\prime})^{2}}{4}\right]\frac{1}{x^{2}}\left[1+\mathcal{O}(x)\right] \notag \\
		    &-\frac{F_g^{\prime\prime}}{2F_g^{\prime}}\frac{1}{x}\left[1+\mathcal{O}(x)\right]. \label{eq:app7}
		\end{align}
		The leading term in Eq.~(\ref{eq:app7}), of order $\mathcal{O}(1/x^2)$, becomes asymptotically dominant  $r\overset{(\mathcal{H})}{\operatorname*{\sim}}r_{g}$.
		Thus, we find the the CQM equation
		\begin{equation}
		    u^{\prime\prime}(x)+\frac{\lambda_{\mathrm{eff}}}{x^2}\left[1+\mathcal{O}(x)\right]u(x)=0, \label{eq:app8}
		\end{equation}
		where we have replaced $u(r)$ by $u(x)$, and
		\begin{equation}
		    \quad\lambda_{\mathrm{eff}}=\frac{1}{4}+\Theta^2,\quad\Theta=\frac{\nu\sqrt{1+\ell}}{F_g^{\prime}}.\label{eq:app9}
		\end{equation}
		The validity of the expansion in $x$ relies on the condition $r-r_g=x\ll r_g$.

\begin{thebibliography}{96}%
\makeatletter
\providecommand \@ifxundefined [1]{%
 \@ifx{#1\undefined}
}%
\providecommand \@ifnum [1]{%
 \ifnum #1\expandafter \@firstoftwo
 \else \expandafter \@secondoftwo
 \fi
}%
\providecommand \@ifx [1]{%
 \ifx #1\expandafter \@firstoftwo
 \else \expandafter \@secondoftwo
 \fi
}%
\providecommand \natexlab [1]{#1}%
\providecommand \enquote  [1]{``#1''}%
\providecommand \bibnamefont  [1]{#1}%
\providecommand \bibfnamefont [1]{#1}%
\providecommand \citenamefont [1]{#1}%
\providecommand \href@noop [0]{\@secondoftwo}%
\providecommand \href [0]{\begingroup \@sanitize@url \@href}%
\providecommand \@href[1]{\@@startlink{#1}\@@href}%
\providecommand \@@href[1]{\endgroup#1\@@endlink}%
\providecommand \@sanitize@url [0]{\catcode `\\12\catcode `\$12\catcode
  `\&12\catcode `\#12\catcode `\^12\catcode `\_12\catcode `\%12\relax}%
\providecommand \@@startlink[1]{}%
\providecommand \@@endlink[0]{}%
\providecommand \url  [0]{\begingroup\@sanitize@url \@url }%
\providecommand \@url [1]{\endgroup\@href {#1}{\urlprefix }}%
\providecommand \urlprefix  [0]{URL }%
\providecommand \Eprint [0]{\href }%
\providecommand \doibase [0]{https://doi.org/}%
\providecommand \selectlanguage [0]{\@gobble}%
\providecommand \bibinfo  [0]{\@secondoftwo}%
\providecommand \bibfield  [0]{\@secondoftwo}%
\providecommand \translation [1]{[#1]}%
\providecommand \BibitemOpen [0]{}%
\providecommand \bibitemStop [0]{}%
\providecommand \bibitemNoStop [0]{.\EOS\space}%
\providecommand \EOS [0]{\spacefactor3000\relax}%
\providecommand \BibitemShut  [1]{\csname bibitem#1\endcsname}%
\let\auto@bib@innerbib\@empty
\bibitem [{\citenamefont {Yagi}\ \emph {et~al.}(2014)\citenamefont {Yagi},
  \citenamefont {Blas}, \citenamefont {Yunes},\ and\ \citenamefont
  {Barausse}}]{yagi2014strong}%
  \BibitemOpen
  \bibfield  {author} {\bibinfo {author} {\bibfnamefont {K.}~\bibnamefont
  {Yagi}}, \bibinfo {author} {\bibfnamefont {D.}~\bibnamefont {Blas}}, \bibinfo
  {author} {\bibfnamefont {N.}~\bibnamefont {Yunes}},\ and\ \bibinfo {author}
  {\bibfnamefont {E.}~\bibnamefont {Barausse}},\ }\bibfield  {title} {\bibinfo
  {title} {{Strong Binary Pulsar Constraints on Lorentz Violation in
  Gravity}},\ }\href {https://doi.org/10.1103/PhysRevLett.112.161101}
  {\bibfield  {journal} {\bibinfo  {journal} {Phys. Rev. Lett.}\ }\textbf
  {\bibinfo {volume} {112}},\ \bibinfo {pages} {161101} (\bibinfo {year}
  {2014})},\ \Eprint {https://arxiv.org/abs/1307.6219} {arXiv:1307.6219
  [gr-qc]} \BibitemShut {NoStop}%
\bibitem [{\citenamefont {Shao}(2014{\natexlab{a}})}]{shao2014tests}%
  \BibitemOpen
  \bibfield  {author} {\bibinfo {author} {\bibfnamefont {L.}~\bibnamefont
  {Shao}},\ }\bibfield  {title} {\bibinfo {title} {{Tests of local Lorentz
  invariance violation of gravity in the standard model extension with
  pulsars}},\ }\href {https://doi.org/10.1103/PhysRevLett.112.111103}
  {\bibfield  {journal} {\bibinfo  {journal} {Phys. Rev. Lett.}\ }\textbf
  {\bibinfo {volume} {112}},\ \bibinfo {pages} {111103} (\bibinfo {year}
  {2014}{\natexlab{a}})},\ \Eprint {https://arxiv.org/abs/1402.6452}
  {arXiv:1402.6452 [gr-qc]} \BibitemShut {NoStop}%
\bibitem [{\citenamefont {Shao}(2014{\natexlab{b}})}]{shao2014new}%
  \BibitemOpen
  \bibfield  {author} {\bibinfo {author} {\bibfnamefont {L.}~\bibnamefont
  {Shao}},\ }\bibfield  {title} {\bibinfo {title} {{New pulsar limit on local
  Lorentz invariance violation of gravity in the standard-model extension}},\
  }\href {https://doi.org/10.1103/PhysRevD.90.122009} {\bibfield  {journal}
  {\bibinfo  {journal} {Phys. Rev. D}\ }\textbf {\bibinfo {volume} {90}},\
  \bibinfo {pages} {122009} (\bibinfo {year} {2014}{\natexlab{b}})},\ \Eprint
  {https://arxiv.org/abs/1412.2320} {arXiv:1412.2320 [gr-qc]} \BibitemShut
  {NoStop}%
\bibitem [{\citenamefont {Perennes}\ \emph {et~al.}(2020)\citenamefont
  {Perennes}, \citenamefont {Sol},\ and\ \citenamefont
  {Bolmont}}]{perennes2020modeling}%
  \BibitemOpen
  \bibfield  {author} {\bibinfo {author} {\bibfnamefont {C.}~\bibnamefont
  {Perennes}}, \bibinfo {author} {\bibfnamefont {H.}~\bibnamefont {Sol}},\ and\
  \bibinfo {author} {\bibfnamefont {J.}~\bibnamefont {Bolmont}},\ }\bibfield
  {title} {\bibinfo {title} {{Modeling spectral lags in active galactic nucleus
  flares in the context of Lorentz invariance violation searches}},\ }\href
  {https://doi.org/10.1051/0004-6361/201936430} {\bibfield  {journal} {\bibinfo
   {journal} {Astron. Astrophys.}\ }\textbf {\bibinfo {volume} {633}},\
  \bibinfo {pages} {A143} (\bibinfo {year} {2020})},\ \Eprint
  {https://arxiv.org/abs/1911.10377} {arXiv:1911.10377 [astro-ph.HE]}
  \BibitemShut {NoStop}%
\bibitem [{\citenamefont {Ellis}\ \emph {et~al.}(2006)\citenamefont {Ellis},
  \citenamefont {Mavromatos}, \citenamefont {Nanopoulos}, \citenamefont
  {Sakharov},\ and\ \citenamefont {Sarkisyan}}]{ellis2006robust}%
  \BibitemOpen
  \bibfield  {author} {\bibinfo {author} {\bibfnamefont {J.~R.}\ \bibnamefont
  {Ellis}}, \bibinfo {author} {\bibfnamefont {N.~E.}\ \bibnamefont
  {Mavromatos}}, \bibinfo {author} {\bibfnamefont {D.~V.}\ \bibnamefont
  {Nanopoulos}}, \bibinfo {author} {\bibfnamefont {A.~S.}\ \bibnamefont
  {Sakharov}},\ and\ \bibinfo {author} {\bibfnamefont {E.~K.~G.}\ \bibnamefont
  {Sarkisyan}},\ }\bibfield  {title} {\bibinfo {title} {{Robust limits on
  Lorentz violation from gamma-ray bursts}},\ }\href
  {https://doi.org/10.1016/j.astropartphys.2007.12.003} {\bibfield  {journal}
  {\bibinfo  {journal} {Astropart. Phys.}\ }\textbf {\bibinfo {volume} {25}},\
  \bibinfo {pages} {402} (\bibinfo {year} {2006})},\ \bibinfo {note} {[Erratum:
  Astropart.Phys. 29, 158--159 (2008)]},\ \Eprint
  {https://arxiv.org/abs/0712.2781} {arXiv:0712.2781 [astro-ph]} \BibitemShut
  {NoStop}%
\bibitem [{\citenamefont {Zhang}\ and\ \citenamefont
  {Ma}(2015)}]{zhang2015lorentz}%
  \BibitemOpen
  \bibfield  {author} {\bibinfo {author} {\bibfnamefont {S.}~\bibnamefont
  {Zhang}}\ and\ \bibinfo {author} {\bibfnamefont {B.-Q.}\ \bibnamefont {Ma}},\
  }\bibfield  {title} {\bibinfo {title} {{Lorentz violation from gamma-ray
  bursts}},\ }\href {https://doi.org/10.1016/j.astropartphys.2014.04.008}
  {\bibfield  {journal} {\bibinfo  {journal} {Astropart. Phys.}\ }\textbf
  {\bibinfo {volume} {61}},\ \bibinfo {pages} {108} (\bibinfo {year} {2015})},\
  \Eprint {https://arxiv.org/abs/1406.4568} {arXiv:1406.4568 [hep-ph]}
  \BibitemShut {NoStop}%
\bibitem [{\citenamefont {Jacob}\ and\ \citenamefont
  {Piran}(2007)}]{jacob2007neutrinos}%
  \BibitemOpen
  \bibfield  {author} {\bibinfo {author} {\bibfnamefont {U.}~\bibnamefont
  {Jacob}}\ and\ \bibinfo {author} {\bibfnamefont {T.}~\bibnamefont {Piran}},\
  }\bibfield  {title} {\bibinfo {title} {{Neutrinos from gamma-ray bursts as a
  tool to explore quantum-gravity-induced Lorentz violation}},\ }\href
  {https://doi.org/10.1038/nphys506} {\bibfield  {journal} {\bibinfo  {journal}
  {Nature Phys.}\ }\textbf {\bibinfo {volume} {3}},\ \bibinfo {pages} {87}
  (\bibinfo {year} {2007})},\ \Eprint {https://arxiv.org/abs/hep-ph/0607145}
  {arXiv:hep-ph/0607145} \BibitemShut {NoStop}%
\bibitem [{\citenamefont {Leon}\ \emph {et~al.}(2016)\citenamefont {Leon},
  \citenamefont {Kaufman}, \citenamefont {Keating},\ and\ \citenamefont
  {Mewes}}]{leon2017cosmic}%
  \BibitemOpen
  \bibfield  {author} {\bibinfo {author} {\bibfnamefont {D.}~\bibnamefont
  {Leon}}, \bibinfo {author} {\bibfnamefont {J.}~\bibnamefont {Kaufman}},
  \bibinfo {author} {\bibfnamefont {B.}~\bibnamefont {Keating}},\ and\ \bibinfo
  {author} {\bibfnamefont {M.}~\bibnamefont {Mewes}},\ }\bibfield  {title}
  {\bibinfo {title} {{The cosmic microwave background and
  pseudo-Nambu\textendash{}Goldstone bosons: Searching for Lorentz violations
  in the cosmos}},\ }\href {https://doi.org/10.1142/S0217732317300026}
  {\bibfield  {journal} {\bibinfo  {journal} {Mod. Phys. Lett. A}\ }\textbf
  {\bibinfo {volume} {32}},\ \bibinfo {pages} {1730002} (\bibinfo {year}
  {2016})},\ \Eprint {https://arxiv.org/abs/1611.00418} {arXiv:1611.00418
  [astro-ph.CO]} \BibitemShut {NoStop}%
\bibitem [{\citenamefont {Kostelecky}\ and\ \citenamefont
  {Samuel}(1989)}]{kostelecky1989phenomenological}%
  \BibitemOpen
  \bibfield  {author} {\bibinfo {author} {\bibfnamefont {V.~A.}\ \bibnamefont
  {Kostelecky}}\ and\ \bibinfo {author} {\bibfnamefont {S.}~\bibnamefont
  {Samuel}},\ }\bibfield  {title} {\bibinfo {title} {{Phenomenological
  Gravitational Constraints on Strings and Higher Dimensional Theories}},\
  }\href {https://doi.org/10.1103/PhysRevLett.63.224} {\bibfield  {journal}
  {\bibinfo  {journal} {Phys. Rev. Lett.}\ }\textbf {\bibinfo {volume} {63}},\
  \bibinfo {pages} {224} (\bibinfo {year} {1989})}\BibitemShut {NoStop}%
\bibitem [{\citenamefont {Carroll}\ \emph {et~al.}(2001)\citenamefont
  {Carroll}, \citenamefont {Harvey}, \citenamefont {Kostelecky}, \citenamefont
  {Lane},\ and\ \citenamefont {Okamoto}}]{Kostelecky2001}%
  \BibitemOpen
  \bibfield  {author} {\bibinfo {author} {\bibfnamefont {S.~M.}\ \bibnamefont
  {Carroll}}, \bibinfo {author} {\bibfnamefont {J.~A.}\ \bibnamefont {Harvey}},
  \bibinfo {author} {\bibfnamefont {V.~A.}\ \bibnamefont {Kostelecky}},
  \bibinfo {author} {\bibfnamefont {C.~D.}\ \bibnamefont {Lane}},\ and\
  \bibinfo {author} {\bibfnamefont {T.}~\bibnamefont {Okamoto}},\ }\bibfield
  {title} {\bibinfo {title} {{Noncommutative field theory and Lorentz
  violation}},\ }\href {https://doi.org/10.1103/PhysRevLett.87.141601}
  {\bibfield  {journal} {\bibinfo  {journal} {Phys. Rev. Lett.}\ }\textbf
  {\bibinfo {volume} {87}},\ \bibinfo {pages} {141601} (\bibinfo {year}
  {2001})},\ \Eprint {https://arxiv.org/abs/0105082} {arXiv:0105082 [hep-th]}
  \BibitemShut {NoStop}%
\bibitem [{\citenamefont {Ashtekar}\ and\ \citenamefont
  {Bianchi}(2021)}]{Ashtekar:2021kfp}%
  \BibitemOpen
  \bibfield  {author} {\bibinfo {author} {\bibfnamefont {A.}~\bibnamefont
  {Ashtekar}}\ and\ \bibinfo {author} {\bibfnamefont {E.}~\bibnamefont
  {Bianchi}},\ }\bibfield  {title} {\bibinfo {title} {{A short review of loop
  quantum gravity}},\ }\href {https://doi.org/10.1088/1361-6633/abed91}
  {\bibfield  {journal} {\bibinfo  {journal} {Rept. Prog. Phys.}\ }\textbf
  {\bibinfo {volume} {84}},\ \bibinfo {pages} {042001} (\bibinfo {year}
  {2021})},\ \Eprint {https://arxiv.org/abs/2104.04394} {arXiv:2104.04394
  [gr-qc]} \BibitemShut {NoStop}%
\bibitem [{\citenamefont {Lewandowski}\ \emph {et~al.}(2023)\citenamefont
  {Lewandowski}, \citenamefont {Ma}, \citenamefont {Yang},\ and\ \citenamefont
  {Zhang}}]{Lewandowski:2022zce}%
  \BibitemOpen
  \bibfield  {author} {\bibinfo {author} {\bibfnamefont {J.}~\bibnamefont
  {Lewandowski}}, \bibinfo {author} {\bibfnamefont {Y.}~\bibnamefont {Ma}},
  \bibinfo {author} {\bibfnamefont {J.}~\bibnamefont {Yang}},\ and\ \bibinfo
  {author} {\bibfnamefont {C.}~\bibnamefont {Zhang}},\ }\bibfield  {title}
  {\bibinfo {title} {{Quantum Oppenheimer-Snyder and Swiss Cheese Models}},\
  }\href {https://doi.org/10.1103/PhysRevLett.130.101501} {\bibfield  {journal}
  {\bibinfo  {journal} {Phys. Rev. Lett.}\ }\textbf {\bibinfo {volume} {130}},\
  \bibinfo {pages} {101501} (\bibinfo {year} {2023})},\ \Eprint
  {https://arxiv.org/abs/2210.02253} {arXiv:2210.02253 [gr-qc]} \BibitemShut
  {NoStop}%
\bibitem [{\citenamefont {Zhang}\ \emph
  {et~al.}(2024{\natexlab{a}})\citenamefont {Zhang}, \citenamefont
  {Lewandowski}, \citenamefont {Ma},\ and\ \citenamefont
  {Yang}}]{Zhang:2024khj}%
  \BibitemOpen
  \bibfield  {author} {\bibinfo {author} {\bibfnamefont {C.}~\bibnamefont
  {Zhang}}, \bibinfo {author} {\bibfnamefont {J.}~\bibnamefont {Lewandowski}},
  \bibinfo {author} {\bibfnamefont {Y.}~\bibnamefont {Ma}},\ and\ \bibinfo
  {author} {\bibfnamefont {J.}~\bibnamefont {Yang}},\ }\bibfield  {title}
  {\bibinfo {title} {{Black Holes and Covariance in Effective Quantum
  Gravity}},\ }\href@noop {} {\  (\bibinfo {year} {2024}{\natexlab{a}})},\
  \Eprint {https://arxiv.org/abs/2407.10168} {arXiv:2407.10168 [gr-qc]}
  \BibitemShut {NoStop}%
\bibitem [{\citenamefont {Zhang}\ \emph
  {et~al.}(2024{\natexlab{b}})\citenamefont {Zhang}, \citenamefont
  {Lewandowski}, \citenamefont {Ma},\ and\ \citenamefont
  {Yang}}]{Zhang:2024ney}%
  \BibitemOpen
  \bibfield  {author} {\bibinfo {author} {\bibfnamefont {C.}~\bibnamefont
  {Zhang}}, \bibinfo {author} {\bibfnamefont {J.}~\bibnamefont {Lewandowski}},
  \bibinfo {author} {\bibfnamefont {Y.}~\bibnamefont {Ma}},\ and\ \bibinfo
  {author} {\bibfnamefont {J.}~\bibnamefont {Yang}},\ }\bibfield  {title}
  {\bibinfo {title} {{Black holes and covariance in effective quantum gravity:
  A solution without Cauchy horizons}},\ }\href@noop {} {\  (\bibinfo {year}
  {2024}{\natexlab{b}})},\ \Eprint {https://arxiv.org/abs/2412.02487}
  {arXiv:2412.02487 [gr-qc]} \BibitemShut {NoStop}%
\bibitem [{\citenamefont {Liu}\ \emph {et~al.}(2024{\natexlab{a}})\citenamefont
  {Liu}, \citenamefont {Wu},\ and\ \citenamefont {Wang}}]{Liu:2024soc}%
  \BibitemOpen
  \bibfield  {author} {\bibinfo {author} {\bibfnamefont {W.}~\bibnamefont
  {Liu}}, \bibinfo {author} {\bibfnamefont {D.}~\bibnamefont {Wu}},\ and\
  \bibinfo {author} {\bibfnamefont {J.}~\bibnamefont {Wang}},\ }\bibfield
  {title} {\bibinfo {title} {{Light rings and shadows of static black holes in
  effective quantum gravity}},\ }\href
  {https://doi.org/10.1016/j.physletb.2024.139052} {\bibfield  {journal}
  {\bibinfo  {journal} {Phys. Lett. B}\ }\textbf {\bibinfo {volume} {858}},\
  \bibinfo {pages} {139052} (\bibinfo {year} {2024}{\natexlab{a}})},\ \Eprint
  {https://arxiv.org/abs/2408.05569} {arXiv:2408.05569 [gr-qc]} \BibitemShut
  {NoStop}%
\bibitem [{\citenamefont {Liu}\ \emph {et~al.}(2024{\natexlab{b}})\citenamefont
  {Liu}, \citenamefont {Wu},\ and\ \citenamefont {Wang}}]{Liu:2024iec}%
  \BibitemOpen
  \bibfield  {author} {\bibinfo {author} {\bibfnamefont {W.}~\bibnamefont
  {Liu}}, \bibinfo {author} {\bibfnamefont {D.}~\bibnamefont {Wu}},\ and\
  \bibinfo {author} {\bibfnamefont {J.}~\bibnamefont {Wang}},\ }\bibfield
  {title} {\bibinfo {title} {{Light rings and shadows of static black holes in
  effective quantum gravity II: A new solution without Cauchy horizons}},\
  }\href@noop {} {\  (\bibinfo {year} {2024}{\natexlab{b}})},\ \Eprint
  {https://arxiv.org/abs/2412.18083} {arXiv:2412.18083 [gr-qc]} \BibitemShut
  {NoStop}%
\bibitem [{\citenamefont {Fu}\ \emph {et~al.}(2024)\citenamefont {Fu},
  \citenamefont {Liu}, \citenamefont {Wang}, \citenamefont {Wu},\ and\
  \citenamefont {Zhang}}]{Fu:2024cfk}%
  \BibitemOpen
  \bibfield  {author} {\bibinfo {author} {\bibfnamefont {G.}~\bibnamefont
  {Fu}}, \bibinfo {author} {\bibfnamefont {Y.}~\bibnamefont {Liu}}, \bibinfo
  {author} {\bibfnamefont {B.}~\bibnamefont {Wang}}, \bibinfo {author}
  {\bibfnamefont {J.-P.}\ \bibnamefont {Wu}},\ and\ \bibinfo {author}
  {\bibfnamefont {C.}~\bibnamefont {Zhang}},\ }\bibfield  {title} {\bibinfo
  {title} {{Probing Quantum Gravity Effects with Eccentric Extreme Mass-Ratio
  Inspirals}},\ }\href@noop {} {\  (\bibinfo {year} {2024})},\ \Eprint
  {https://arxiv.org/abs/2409.08138} {arXiv:2409.08138 [gr-qc]} \BibitemShut
  {NoStop}%
\bibitem [{\citenamefont {Choi}\ and\ \citenamefont
  {Park}(2015)}]{choi2015standard}%
  \BibitemOpen
  \bibfield  {author} {\bibinfo {author} {\bibfnamefont {K.-S.}\ \bibnamefont
  {Choi}}\ and\ \bibinfo {author} {\bibfnamefont {J.-H.}\ \bibnamefont
  {Park}},\ }\bibfield  {title} {\bibinfo {title} {{Standard Model as a Double
  Field Theory}},\ }\href {https://doi.org/10.1103/PhysRevLett.115.171603}
  {\bibfield  {journal} {\bibinfo  {journal} {Phys. Rev. Lett.}\ }\textbf
  {\bibinfo {volume} {115}},\ \bibinfo {pages} {171603} (\bibinfo {year}
  {2015})},\ \Eprint {https://arxiv.org/abs/1506.05277} {arXiv:1506.05277
  [hep-th]} \BibitemShut {NoStop}%
\bibitem [{\citenamefont {Del~Porro}\ \emph {et~al.}(2023)\citenamefont
  {Del~Porro}, \citenamefont {Herrero-Valea}, \citenamefont {Liberati},\ and\
  \citenamefont {Schneider}}]{del2023hawking}%
  \BibitemOpen
  \bibfield  {author} {\bibinfo {author} {\bibfnamefont {F.}~\bibnamefont
  {Del~Porro}}, \bibinfo {author} {\bibfnamefont {M.}~\bibnamefont
  {Herrero-Valea}}, \bibinfo {author} {\bibfnamefont {S.}~\bibnamefont
  {Liberati}},\ and\ \bibinfo {author} {\bibfnamefont {M.}~\bibnamefont
  {Schneider}},\ }\bibfield  {title} {\bibinfo {title} {{Hawking radiation in
  Lorentz violating gravity: a tale of two horizons}},\ }\href
  {https://doi.org/10.1007/JHEP12(2023)094} {\bibfield  {journal} {\bibinfo
  {journal} {JHEP}\ }\textbf {\bibinfo {volume} {12}},\ \bibinfo {pages}
  {094}},\ \Eprint {https://arxiv.org/abs/2310.01472} {arXiv:2310.01472
  [gr-qc]} \BibitemShut {NoStop}%
\bibitem [{\citenamefont {Tian}\ \emph {et~al.}(2022)\citenamefont {Tian},
  \citenamefont {Wu}, \citenamefont {Zhang}, \citenamefont {Jing},\ and\
  \citenamefont {Du}}]{tian2022probing}%
  \BibitemOpen
  \bibfield  {author} {\bibinfo {author} {\bibfnamefont {Z.}~\bibnamefont
  {Tian}}, \bibinfo {author} {\bibfnamefont {L.}~\bibnamefont {Wu}}, \bibinfo
  {author} {\bibfnamefont {L.}~\bibnamefont {Zhang}}, \bibinfo {author}
  {\bibfnamefont {J.}~\bibnamefont {Jing}},\ and\ \bibinfo {author}
  {\bibfnamefont {J.}~\bibnamefont {Du}},\ }\bibfield  {title} {\bibinfo
  {title} {{Probing Lorentz-invariance-violation-induced nonthermal Unruh
  effect in quasi-two-dimensional dipolar condensates}},\ }\href
  {https://doi.org/10.1103/PhysRevD.106.L061701} {\bibfield  {journal}
  {\bibinfo  {journal} {Phys. Rev. D}\ }\textbf {\bibinfo {volume} {106}},\
  \bibinfo {pages} {L061701} (\bibinfo {year} {2022})},\ \Eprint
  {https://arxiv.org/abs/2205.08669} {arXiv:2205.08669 [quant-ph]} \BibitemShut
  {NoStop}%
\bibitem [{\citenamefont {Agarwalla}\ \emph {et~al.}(2023)\citenamefont
  {Agarwalla}, \citenamefont {Das}, \citenamefont {Sahoo},\ and\ \citenamefont
  {Swain}}]{agarwalla2023constraining}%
  \BibitemOpen
  \bibfield  {author} {\bibinfo {author} {\bibfnamefont {S.~K.}\ \bibnamefont
  {Agarwalla}}, \bibinfo {author} {\bibfnamefont {S.}~\bibnamefont {Das}},
  \bibinfo {author} {\bibfnamefont {S.}~\bibnamefont {Sahoo}},\ and\ \bibinfo
  {author} {\bibfnamefont {P.}~\bibnamefont {Swain}},\ }\bibfield  {title}
  {\bibinfo {title} {{Constraining Lorentz invariance violation with
  next-generation long-baseline experiments}},\ }\href
  {https://doi.org/10.1007/JHEP07(2023)216} {\bibfield  {journal} {\bibinfo
  {journal} {JHEP}\ }\textbf {\bibinfo {volume} {07}},\ \bibinfo {pages}
  {216}},\ \Eprint {https://arxiv.org/abs/2302.12005} {arXiv:2302.12005
  [hep-ph]} \BibitemShut {NoStop}%
\bibitem [{\citenamefont {Filho}\ \emph
  {et~al.}(2024{\natexlab{a}})\citenamefont {Filho}, \citenamefont {Reis},\
  and\ \citenamefont {Hassanabadi}}]{Filho:2023ycx}%
  \BibitemOpen
  \bibfield  {author} {\bibinfo {author} {\bibfnamefont {A.~A.~A.}\
  \bibnamefont {Filho}}, \bibinfo {author} {\bibfnamefont {J.~A. A.~S.}\
  \bibnamefont {Reis}},\ and\ \bibinfo {author} {\bibfnamefont
  {H.}~\bibnamefont {Hassanabadi}},\ }\bibfield  {title} {\bibinfo {title}
  {{Exploring antisymmetric tensor effects on black hole shadows and
  quasinormal frequencies}},\ }\href
  {https://doi.org/10.1088/1475-7516/2024/05/029} {\bibfield  {journal}
  {\bibinfo  {journal} {JCAP}\ }\textbf {\bibinfo {volume} {05}},\ \bibinfo
  {pages} {029}},\ \Eprint {https://arxiv.org/abs/2309.15778} {arXiv:2309.15778
  [gr-qc]} \BibitemShut {NoStop}%
\bibitem [{\citenamefont {Guo}\ \emph {et~al.}(2024)\citenamefont {Guo},
  \citenamefont {Tan},\ and\ \citenamefont {Liu}}]{Guo:2023nkd}%
  \BibitemOpen
  \bibfield  {author} {\bibinfo {author} {\bibfnamefont {W.-D.}\ \bibnamefont
  {Guo}}, \bibinfo {author} {\bibfnamefont {Q.}~\bibnamefont {Tan}},\ and\
  \bibinfo {author} {\bibfnamefont {Y.-X.}\ \bibnamefont {Liu}},\ }\bibfield
  {title} {\bibinfo {title} {{Quasinormal modes and greybody factor of a
  Lorentz-violating black hole}},\ }\href
  {https://doi.org/10.1088/1475-7516/2024/07/008} {\bibfield  {journal}
  {\bibinfo  {journal} {JCAP}\ }\textbf {\bibinfo {volume} {07}},\ \bibinfo
  {pages} {008}},\ \Eprint {https://arxiv.org/abs/2312.16605} {arXiv:2312.16605
  [gr-qc]} \BibitemShut {NoStop}%
\bibitem [{\citenamefont {Filho}\ \emph
  {et~al.}(2024{\natexlab{b}})\citenamefont {Filho}, \citenamefont {Heidari},
  \citenamefont {Reis},\ and\ \citenamefont {Hassanabadi}}]{Filho:2024kbq}%
  \BibitemOpen
  \bibfield  {author} {\bibinfo {author} {\bibfnamefont {A.~A.~A.}\
  \bibnamefont {Filho}}, \bibinfo {author} {\bibfnamefont {N.}~\bibnamefont
  {Heidari}}, \bibinfo {author} {\bibfnamefont {J.~A. A.~S.}\ \bibnamefont
  {Reis}},\ and\ \bibinfo {author} {\bibfnamefont {H.}~\bibnamefont
  {Hassanabadi}},\ }\bibfield  {title} {\bibinfo {title} {{The impact of an
  antisymmetric tensor on charged black holes: evaporation process, geodesics,
  deflection angle, scattering effects and quasinormal modes}},\ }\href@noop {}
  {\  (\bibinfo {year} {2024}{\natexlab{b}})},\ \Eprint
  {https://arxiv.org/abs/2404.10721} {arXiv:2404.10721 [gr-qc]} \BibitemShut
  {NoStop}%
\bibitem [{\citenamefont {Ara\'ujo~Filho}\ \emph {et~al.}(2024)\citenamefont
  {Ara\'ujo~Filho}, \citenamefont {Heidari}, \citenamefont {Reis},\ and\
  \citenamefont {Hassanabadi}}]{AraujoFilho:2024rcr}%
  \BibitemOpen
  \bibfield  {author} {\bibinfo {author} {\bibfnamefont {A.~A.}\ \bibnamefont
  {Ara\'ujo~Filho}}, \bibinfo {author} {\bibfnamefont {N.}~\bibnamefont
  {Heidari}}, \bibinfo {author} {\bibfnamefont {J.~A. A.~S.}\ \bibnamefont
  {Reis}},\ and\ \bibinfo {author} {\bibfnamefont {H.}~\bibnamefont
  {Hassanabadi}},\ }\bibfield  {title} {\bibinfo {title} {{The impact of an
  antisymmetric tensor on charged black holes: evaporation process, geodesics,
  deflection angle, scattering effects and quasinormal modes}},\ }\href@noop {}
  {\  (\bibinfo {year} {2024})},\ \Eprint {https://arxiv.org/abs/2404.10721}
  {arXiv:2404.10721 [gr-qc]} \BibitemShut {NoStop}%
\bibitem [{\citenamefont {Filho}(2024)}]{Filho:2024tgy}%
  \BibitemOpen
  \bibfield  {author} {\bibinfo {author} {\bibfnamefont {A.~A.~A.}\
  \bibnamefont {Filho}},\ }\bibfield  {title} {\bibinfo {title} {{Antisymmetric
  tensor influence on charged black hole lensing phenomena and time delay}},\
  }\href@noop {} {\  (\bibinfo {year} {2024})},\ \Eprint
  {https://arxiv.org/abs/2406.11582} {arXiv:2406.11582 [gr-qc]} \BibitemShut
  {NoStop}%
\bibitem [{\citenamefont {Hosseinifar}\ \emph {et~al.}(2024)\citenamefont
  {Hosseinifar}, \citenamefont {Filho}, \citenamefont {Zhang}, \citenamefont
  {Chen},\ and\ \citenamefont {Hassanabadi}}]{Hosseinifar:2024wwe}%
  \BibitemOpen
  \bibfield  {author} {\bibinfo {author} {\bibfnamefont {F.}~\bibnamefont
  {Hosseinifar}}, \bibinfo {author} {\bibfnamefont {A.~A.~A.}\ \bibnamefont
  {Filho}}, \bibinfo {author} {\bibfnamefont {M.~Y.}\ \bibnamefont {Zhang}},
  \bibinfo {author} {\bibfnamefont {H.}~\bibnamefont {Chen}},\ and\ \bibinfo
  {author} {\bibfnamefont {H.}~\bibnamefont {Hassanabadi}},\ }\bibfield
  {title} {\bibinfo {title} {{Shadows, greybody factors, emission rate,
  topological charge, and phase transitions for a charged black hole with a
  Kalb-Ramond field background}},\ }\href@noop {} {\  (\bibinfo {year}
  {2024})},\ \Eprint {https://arxiv.org/abs/2407.07017} {arXiv:2407.07017
  [gr-qc]} \BibitemShut {NoStop}%
\bibitem [{\citenamefont {Kalb}\ and\ \citenamefont
  {Ramond}(1974)}]{kalb1974classical}%
  \BibitemOpen
  \bibfield  {author} {\bibinfo {author} {\bibfnamefont {M.}~\bibnamefont
  {Kalb}}\ and\ \bibinfo {author} {\bibfnamefont {P.}~\bibnamefont {Ramond}},\
  }\bibfield  {title} {\bibinfo {title} {{Classical direct interstring
  action}},\ }\href {https://doi.org/10.1103/PhysRevD.9.2273} {\bibfield
  {journal} {\bibinfo  {journal} {Phys. Rev. D}\ }\textbf {\bibinfo {volume}
  {9}},\ \bibinfo {pages} {2273} (\bibinfo {year} {1974})}\BibitemShut
  {NoStop}%
\bibitem [{\citenamefont {Yang}\ \emph {et~al.}(2023)\citenamefont {Yang},
  \citenamefont {Chen}, \citenamefont {Duan},\ and\ \citenamefont
  {Zhao}}]{yang2023static}%
  \BibitemOpen
  \bibfield  {author} {\bibinfo {author} {\bibfnamefont {K.}~\bibnamefont
  {Yang}}, \bibinfo {author} {\bibfnamefont {Y.-Z.}\ \bibnamefont {Chen}},
  \bibinfo {author} {\bibfnamefont {Z.-Q.}\ \bibnamefont {Duan}},\ and\
  \bibinfo {author} {\bibfnamefont {J.-Y.}\ \bibnamefont {Zhao}},\ }\bibfield
  {title} {\bibinfo {title} {{Static and spherically symmetric black holes in
  gravity with a background Kalb-Ramond field}},\ }\href
  {https://doi.org/10.1103/PhysRevD.108.124004} {\bibfield  {journal} {\bibinfo
   {journal} {Phys. Rev. D}\ }\textbf {\bibinfo {volume} {108}},\ \bibinfo
  {pages} {124004} (\bibinfo {year} {2023})},\ \Eprint
  {https://arxiv.org/abs/2308.06613} {arXiv:2308.06613 [gr-qc]} \BibitemShut
  {NoStop}%
\bibitem [{\citenamefont {Duan}\ \emph {et~al.}(2024)\citenamefont {Duan},
  \citenamefont {Zhao},\ and\ \citenamefont {Yang}}]{duan2023electrically}%
  \BibitemOpen
  \bibfield  {author} {\bibinfo {author} {\bibfnamefont {Z.-Q.}\ \bibnamefont
  {Duan}}, \bibinfo {author} {\bibfnamefont {J.-Y.}\ \bibnamefont {Zhao}},\
  and\ \bibinfo {author} {\bibfnamefont {K.}~\bibnamefont {Yang}},\ }\bibfield
  {title} {\bibinfo {title} {{Electrically charged black holes in gravity with
  a background Kalb\textendash{}Ramond field}},\ }\href
  {https://doi.org/10.1140/epjc/s10052-024-13188-5} {\bibfield  {journal}
  {\bibinfo  {journal} {Eur. Phys. J. C}\ }\textbf {\bibinfo {volume} {84}},\
  \bibinfo {pages} {798} (\bibinfo {year} {2024})},\ \Eprint
  {https://arxiv.org/abs/2310.13555} {arXiv:2310.13555 [gr-qc]} \BibitemShut
  {NoStop}%
\bibitem [{\citenamefont {Liu}\ \emph {et~al.}(2024{\natexlab{c}})\citenamefont
  {Liu}, \citenamefont {Wu},\ and\ \citenamefont {Wang}}]{Liu:2024oas}%
  \BibitemOpen
  \bibfield  {author} {\bibinfo {author} {\bibfnamefont {W.}~\bibnamefont
  {Liu}}, \bibinfo {author} {\bibfnamefont {D.}~\bibnamefont {Wu}},\ and\
  \bibinfo {author} {\bibfnamefont {J.}~\bibnamefont {Wang}},\ }\bibfield
  {title} {\bibinfo {title} {{Static neutral black holes in Kalb-Ramond
  gravity}},\ }\href {https://doi.org/10.1088/1475-7516/2024/09/017} {\bibfield
   {journal} {\bibinfo  {journal} {JCAP}\ }\textbf {\bibinfo {volume} {09}},\
  \bibinfo {pages} {017}},\ \Eprint {https://arxiv.org/abs/2406.13461}
  {arXiv:2406.13461 [hep-th]} \BibitemShut {NoStop}%
\bibitem [{\citenamefont {Liu}\ \emph {et~al.}(2024{\natexlab{d}})\citenamefont
  {Liu}, \citenamefont {Wu},\ and\ \citenamefont {Wang}}]{Liu:2024lve}%
  \BibitemOpen
  \bibfield  {author} {\bibinfo {author} {\bibfnamefont {W.}~\bibnamefont
  {Liu}}, \bibinfo {author} {\bibfnamefont {D.}~\bibnamefont {Wu}},\ and\
  \bibinfo {author} {\bibfnamefont {J.}~\bibnamefont {Wang}},\ }\bibfield
  {title} {\bibinfo {title} {{Shadow of slowly rotating Kalb-Ramond black
  holes}},\ }\href@noop {} {\  (\bibinfo {year} {2024}{\natexlab{d}})},\
  \Eprint {https://arxiv.org/abs/2407.07416} {arXiv:2407.07416 [gr-qc]}
  \BibitemShut {NoStop}%
\bibitem [{\citenamefont {Kostelecky}(2004)}]{kostelecky2004gravity}%
  \BibitemOpen
  \bibfield  {author} {\bibinfo {author} {\bibfnamefont {V.~A.}\ \bibnamefont
  {Kostelecky}},\ }\bibfield  {title} {\bibinfo {title} {{Gravity, Lorentz
  violation, and the standard model}},\ }\href
  {https://doi.org/10.1103/PhysRevD.69.105009} {\bibfield  {journal} {\bibinfo
  {journal} {Phys. Rev. D}\ }\textbf {\bibinfo {volume} {69}},\ \bibinfo
  {pages} {105009} (\bibinfo {year} {2004})},\ \Eprint
  {https://arxiv.org/abs/hep-th/0312310} {arXiv:hep-th/0312310} \BibitemShut
  {NoStop}%
\bibitem [{\citenamefont {Casana}\ \emph {et~al.}(2018)\citenamefont {Casana},
  \citenamefont {Cavalcante}, \citenamefont {Poulis},\ and\ \citenamefont
  {Santos}}]{casana2018exact}%
  \BibitemOpen
  \bibfield  {author} {\bibinfo {author} {\bibfnamefont {R.}~\bibnamefont
  {Casana}}, \bibinfo {author} {\bibfnamefont {A.}~\bibnamefont {Cavalcante}},
  \bibinfo {author} {\bibfnamefont {F.~P.}\ \bibnamefont {Poulis}},\ and\
  \bibinfo {author} {\bibfnamefont {E.~B.}\ \bibnamefont {Santos}},\ }\bibfield
   {title} {\bibinfo {title} {{Exact Schwarzschild-like solution in a bumblebee
  gravity model}},\ }\href {https://doi.org/10.1103/PhysRevD.97.104001}
  {\bibfield  {journal} {\bibinfo  {journal} {Phys. Rev. D}\ }\textbf {\bibinfo
  {volume} {97}},\ \bibinfo {pages} {104001} (\bibinfo {year} {2018})},\
  \Eprint {https://arxiv.org/abs/1711.02273} {arXiv:1711.02273 [gr-qc]}
  \BibitemShut {NoStop}%
\bibitem [{\citenamefont {\"Ovg\"un}\ \emph {et~al.}(2019)\citenamefont
  {\"Ovg\"un}, \citenamefont {Jusufi},\ and\ \citenamefont
  {Sakall\i{}}}]{ovgun2019exact}%
  \BibitemOpen
  \bibfield  {author} {\bibinfo {author} {\bibfnamefont {A.}~\bibnamefont
  {\"Ovg\"un}}, \bibinfo {author} {\bibfnamefont {K.}~\bibnamefont {Jusufi}},\
  and\ \bibinfo {author} {\bibfnamefont {I.}~\bibnamefont {Sakall\i{}}},\
  }\bibfield  {title} {\bibinfo {title} {{Exact traversable wormhole solution
  in bumblebee gravity}},\ }\href {https://doi.org/10.1103/PhysRevD.99.024042}
  {\bibfield  {journal} {\bibinfo  {journal} {Phys. Rev. D}\ }\textbf {\bibinfo
  {volume} {99}},\ \bibinfo {pages} {024042} (\bibinfo {year} {2019})},\
  \Eprint {https://arxiv.org/abs/1804.09911} {arXiv:1804.09911 [gr-qc]}
  \BibitemShut {NoStop}%
\bibitem [{\citenamefont {G\"ull\"u}\ and\ \citenamefont
  {\"Ovg\"un}(2022)}]{gullu2022schwarzschild}%
  \BibitemOpen
  \bibfield  {author} {\bibinfo {author} {\bibfnamefont {I.}~\bibnamefont
  {G\"ull\"u}}\ and\ \bibinfo {author} {\bibfnamefont {A.}~\bibnamefont
  {\"Ovg\"un}},\ }\bibfield  {title} {\bibinfo {title} {{Schwarzschild-like
  black hole with a topological defect in bumblebee gravity}},\ }\href
  {https://doi.org/10.1016/j.aop.2021.168721} {\bibfield  {journal} {\bibinfo
  {journal} {Annals Phys.}\ }\textbf {\bibinfo {volume} {436}},\ \bibinfo
  {pages} {168721} (\bibinfo {year} {2022})},\ \Eprint
  {https://arxiv.org/abs/2012.02611} {arXiv:2012.02611 [gr-qc]} \BibitemShut
  {NoStop}%
\bibitem [{\citenamefont {Zhang}\ \emph {et~al.}(2022)\citenamefont {Zhang},
  \citenamefont {Liu}, \citenamefont {Liu},\ and\ \citenamefont
  {Yang}}]{poulis2022exact}%
  \BibitemOpen
  \bibfield  {author} {\bibinfo {author} {\bibfnamefont {J.}~\bibnamefont
  {Zhang}}, \bibinfo {author} {\bibfnamefont {M.}~\bibnamefont {Liu}}, \bibinfo
  {author} {\bibfnamefont {Z.}~\bibnamefont {Liu}},\ and\ \bibinfo {author}
  {\bibfnamefont {S.}~\bibnamefont {Yang}},\ }\bibfield  {title} {\bibinfo
  {title} {{A new touch temperature of the event horizon and Rindler horizon in
  the Kinnersley spacetime}},\ }\href
  {https://doi.org/10.1140/epjc/s10052-021-09951-7} {\bibfield  {journal}
  {\bibinfo  {journal} {Eur. Phys. J. C}\ }\textbf {\bibinfo {volume} {82}},\
  \bibinfo {pages} {1} (\bibinfo {year} {2022})}\BibitemShut {NoStop}%
\bibitem [{\citenamefont {Maluf}\ and\ \citenamefont
  {Neves}(2021)}]{maluf2021black}%
  \BibitemOpen
  \bibfield  {author} {\bibinfo {author} {\bibfnamefont {R.~V.}\ \bibnamefont
  {Maluf}}\ and\ \bibinfo {author} {\bibfnamefont {J.~C.~S.}\ \bibnamefont
  {Neves}},\ }\bibfield  {title} {\bibinfo {title} {{Black holes with a
  cosmological constant in bumblebee gravity}},\ }\href
  {https://doi.org/10.1103/PhysRevD.103.044002} {\bibfield  {journal} {\bibinfo
   {journal} {Phys. Rev. D}\ }\textbf {\bibinfo {volume} {103}},\ \bibinfo
  {pages} {044002} (\bibinfo {year} {2021})},\ \Eprint
  {https://arxiv.org/abs/2011.12841} {arXiv:2011.12841 [gr-qc]} \BibitemShut
  {NoStop}%
\bibitem [{\citenamefont {Xu}\ \emph {et~al.}(2023{\natexlab{a}})\citenamefont
  {Xu}, \citenamefont {Liang},\ and\ \citenamefont {Shao}}]{xu2023static}%
  \BibitemOpen
  \bibfield  {author} {\bibinfo {author} {\bibfnamefont {R.}~\bibnamefont
  {Xu}}, \bibinfo {author} {\bibfnamefont {D.}~\bibnamefont {Liang}},\ and\
  \bibinfo {author} {\bibfnamefont {L.}~\bibnamefont {Shao}},\ }\bibfield
  {title} {\bibinfo {title} {{Static spherical vacuum solutions in the
  bumblebee gravity model}},\ }\href
  {https://doi.org/10.1103/PhysRevD.107.024011} {\bibfield  {journal} {\bibinfo
   {journal} {Phys. Rev. D}\ }\textbf {\bibinfo {volume} {107}},\ \bibinfo
  {pages} {024011} (\bibinfo {year} {2023}{\natexlab{a}})},\ \Eprint
  {https://arxiv.org/abs/2209.02209} {arXiv:2209.02209 [gr-qc]} \BibitemShut
  {NoStop}%
\bibitem [{\citenamefont {Ding}\ \emph {et~al.}(2022)\citenamefont {Ding},
  \citenamefont {Chen},\ and\ \citenamefont {Fu}}]{ding2022einstein}%
  \BibitemOpen
  \bibfield  {author} {\bibinfo {author} {\bibfnamefont {C.}~\bibnamefont
  {Ding}}, \bibinfo {author} {\bibfnamefont {X.}~\bibnamefont {Chen}},\ and\
  \bibinfo {author} {\bibfnamefont {X.}~\bibnamefont {Fu}},\ }\bibfield
  {title} {\bibinfo {title} {{Einstein-Gauss-Bonnet gravity coupled to
  bumblebee field in four dimensional spacetime}},\ }\href
  {https://doi.org/10.1016/j.nuclphysb.2022.115688} {\bibfield  {journal}
  {\bibinfo  {journal} {Nucl. Phys. B}\ }\textbf {\bibinfo {volume} {975}},\
  \bibinfo {pages} {115688} (\bibinfo {year} {2022})},\ \Eprint
  {https://arxiv.org/abs/2102.13335} {arXiv:2102.13335 [gr-qc]} \BibitemShut
  {NoStop}%
\bibitem [{\citenamefont {Liu}\ \emph {et~al.}(2024{\natexlab{e}})\citenamefont
  {Liu}, \citenamefont {Guo}, \citenamefont {Wei},\ and\ \citenamefont
  {Liu}}]{Liu:2024axg}%
  \BibitemOpen
  \bibfield  {author} {\bibinfo {author} {\bibfnamefont {J.-Z.}\ \bibnamefont
  {Liu}}, \bibinfo {author} {\bibfnamefont {W.-D.}\ \bibnamefont {Guo}},
  \bibinfo {author} {\bibfnamefont {S.-W.}\ \bibnamefont {Wei}},\ and\ \bibinfo
  {author} {\bibfnamefont {Y.-X.}\ \bibnamefont {Liu}},\ }\bibfield  {title}
  {\bibinfo {title} {{Charged spherically symmetric and slowly rotating charged
  black hole solutions in bumblebee gravity}},\ }\href@noop {} {\  (\bibinfo
  {year} {2024}{\natexlab{e}})},\ \Eprint {https://arxiv.org/abs/2407.08396}
  {arXiv:2407.08396 [gr-qc]} \BibitemShut {NoStop}%
\bibitem [{\citenamefont {Liu}\ \emph {et~al.}(2023)\citenamefont {Liu},
  \citenamefont {Fang}, \citenamefont {Jing},\ and\ \citenamefont
  {Wang}}]{Liu:2022dcn}%
  \BibitemOpen
  \bibfield  {author} {\bibinfo {author} {\bibfnamefont {W.}~\bibnamefont
  {Liu}}, \bibinfo {author} {\bibfnamefont {X.}~\bibnamefont {Fang}}, \bibinfo
  {author} {\bibfnamefont {J.}~\bibnamefont {Jing}},\ and\ \bibinfo {author}
  {\bibfnamefont {J.}~\bibnamefont {Wang}},\ }\bibfield  {title} {\bibinfo
  {title} {{QNMs of slowly rotating Einstein\textendash{}Bumblebee black
  hole}},\ }\href {https://doi.org/10.1140/epjc/s10052-023-11231-5} {\bibfield
  {journal} {\bibinfo  {journal} {Eur. Phys. J. C}\ }\textbf {\bibinfo {volume}
  {83}},\ \bibinfo {pages} {83} (\bibinfo {year} {2023})},\ \Eprint
  {https://arxiv.org/abs/2211.03156} {arXiv:2211.03156 [gr-qc]} \BibitemShut
  {NoStop}%
\bibitem [{\citenamefont {Mai}\ \emph {et~al.}(2023)\citenamefont {Mai},
  \citenamefont {Xu}, \citenamefont {Liang},\ and\ \citenamefont
  {Shao}}]{mai2023extended}%
  \BibitemOpen
  \bibfield  {author} {\bibinfo {author} {\bibfnamefont {Z.-F.}\ \bibnamefont
  {Mai}}, \bibinfo {author} {\bibfnamefont {R.}~\bibnamefont {Xu}}, \bibinfo
  {author} {\bibfnamefont {D.}~\bibnamefont {Liang}},\ and\ \bibinfo {author}
  {\bibfnamefont {L.}~\bibnamefont {Shao}},\ }\bibfield  {title} {\bibinfo
  {title} {{Extended thermodynamics of the bumblebee black holes}},\ }\href
  {https://doi.org/10.1103/PhysRevD.108.024004} {\bibfield  {journal} {\bibinfo
   {journal} {Phys. Rev. D}\ }\textbf {\bibinfo {volume} {108}},\ \bibinfo
  {pages} {024004} (\bibinfo {year} {2023})},\ \Eprint
  {https://arxiv.org/abs/2304.08030} {arXiv:2304.08030 [gr-qc]} \BibitemShut
  {NoStop}%
\bibitem [{\citenamefont {Xu}\ \emph {et~al.}(2023{\natexlab{b}})\citenamefont
  {Xu}, \citenamefont {Liang},\ and\ \citenamefont {Shao}}]{xu2023bumblebee}%
  \BibitemOpen
  \bibfield  {author} {\bibinfo {author} {\bibfnamefont {R.}~\bibnamefont
  {Xu}}, \bibinfo {author} {\bibfnamefont {D.}~\bibnamefont {Liang}},\ and\
  \bibinfo {author} {\bibfnamefont {L.}~\bibnamefont {Shao}},\ }\bibfield
  {title} {\bibinfo {title} {{Bumblebee Black Holes in Light of Event Horizon
  Telescope Observations}},\ }\href {https://doi.org/10.3847/1538-4357/acbdfb}
  {\bibfield  {journal} {\bibinfo  {journal} {Astrophys. J.}\ }\textbf
  {\bibinfo {volume} {945}},\ \bibinfo {pages} {148} (\bibinfo {year}
  {2023}{\natexlab{b}})},\ \Eprint {https://arxiv.org/abs/2302.05671}
  {arXiv:2302.05671 [gr-qc]} \BibitemShut {NoStop}%
\bibitem [{\citenamefont {Zhang}\ \emph {et~al.}(2023)\citenamefont {Zhang},
  \citenamefont {Wang},\ and\ \citenamefont {Jing}}]{zhang2023quasinormal}%
  \BibitemOpen
  \bibfield  {author} {\bibinfo {author} {\bibfnamefont {X.}~\bibnamefont
  {Zhang}}, \bibinfo {author} {\bibfnamefont {M.}~\bibnamefont {Wang}},\ and\
  \bibinfo {author} {\bibfnamefont {J.}~\bibnamefont {Jing}},\ }\bibfield
  {title} {\bibinfo {title} {{Quasinormal modes and late time tails of
  perturbation fields on a Schwarzschild-like black hole with a global monopole
  in the Einstein-bumblebee theory}},\ }\href
  {https://doi.org/10.1007/s11433-023-2153-6} {\bibfield  {journal} {\bibinfo
  {journal} {Sci. China Phys. Mech. Astron.}\ }\textbf {\bibinfo {volume}
  {66}},\ \bibinfo {pages} {100411} (\bibinfo {year} {2023})},\ \Eprint
  {https://arxiv.org/abs/2307.10856} {arXiv:2307.10856 [gr-qc]} \BibitemShut
  {NoStop}%
\bibitem [{\citenamefont {Lin}\ \emph {et~al.}(2023)\citenamefont {Lin},
  \citenamefont {Jiang},\ and\ \citenamefont {Zhai}}]{lin2023quasinormal}%
  \BibitemOpen
  \bibfield  {author} {\bibinfo {author} {\bibfnamefont {R.-H.}\ \bibnamefont
  {Lin}}, \bibinfo {author} {\bibfnamefont {R.}~\bibnamefont {Jiang}},\ and\
  \bibinfo {author} {\bibfnamefont {X.-H.}\ \bibnamefont {Zhai}},\ }\bibfield
  {title} {\bibinfo {title} {{Quasinormal modes of the spherical bumblebee
  black holes with a global monopole}},\ }\href
  {https://doi.org/10.1140/epjc/s10052-023-11899-9} {\bibfield  {journal}
  {\bibinfo  {journal} {Eur. Phys. J. C}\ }\textbf {\bibinfo {volume} {83}},\
  \bibinfo {pages} {720} (\bibinfo {year} {2023})},\ \Eprint
  {https://arxiv.org/abs/2308.01575} {arXiv:2308.01575 [gr-qc]} \BibitemShut
  {NoStop}%
\bibitem [{\citenamefont {Chen}\ \emph
  {et~al.}(2023{\natexlab{a}})\citenamefont {Chen}, \citenamefont {Pan},\ and\
  \citenamefont {Jing}}]{chen2023quasinormal}%
  \BibitemOpen
  \bibfield  {author} {\bibinfo {author} {\bibfnamefont {C.}~\bibnamefont
  {Chen}}, \bibinfo {author} {\bibfnamefont {Q.}~\bibnamefont {Pan}},\ and\
  \bibinfo {author} {\bibfnamefont {J.}~\bibnamefont {Jing}},\ }\bibfield
  {title} {\bibinfo {title} {{Quasinormal modes of a scalar perturbation around
  a rotating BTZ-like black hole in Einstein-bumblebee gravity}},\ }\href
  {https://doi.org/10.1016/j.physletb.2023.138186} {\bibfield  {journal}
  {\bibinfo  {journal} {Phys. Lett. B}\ }\textbf {\bibinfo {volume} {846}},\
  \bibinfo {pages} {138186} (\bibinfo {year} {2023}{\natexlab{a}})},\ \Eprint
  {https://arxiv.org/abs/2302.05861} {arXiv:2302.05861 [gr-qc]} \BibitemShut
  {NoStop}%
\bibitem [{\citenamefont {Chen}\ \emph {et~al.}(2020)\citenamefont {Chen},
  \citenamefont {Wang},\ and\ \citenamefont {Jing}}]{chen2020polarization}%
  \BibitemOpen
  \bibfield  {author} {\bibinfo {author} {\bibfnamefont {S.}~\bibnamefont
  {Chen}}, \bibinfo {author} {\bibfnamefont {M.}~\bibnamefont {Wang}},\ and\
  \bibinfo {author} {\bibfnamefont {J.}~\bibnamefont {Jing}},\ }\bibfield
  {title} {\bibinfo {title} {{Polarization effects in Kerr black hole shadow
  due to the coupling between photon and bumblebee field}},\ }\href
  {https://doi.org/10.1007/JHEP07(2020)054} {\bibfield  {journal} {\bibinfo
  {journal} {JHEP}\ }\textbf {\bibinfo {volume} {07}},\ \bibinfo {pages}
  {054}},\ \Eprint {https://arxiv.org/abs/2004.08857} {arXiv:2004.08857
  [gr-qc]} \BibitemShut {NoStop}%
\bibitem [{\citenamefont {Wang}\ \emph {et~al.}(2022)\citenamefont {Wang},
  \citenamefont {Chen},\ and\ \citenamefont {Jing}}]{wang2022constraint}%
  \BibitemOpen
  \bibfield  {author} {\bibinfo {author} {\bibfnamefont {Z.}~\bibnamefont
  {Wang}}, \bibinfo {author} {\bibfnamefont {S.}~\bibnamefont {Chen}},\ and\
  \bibinfo {author} {\bibfnamefont {J.}~\bibnamefont {Jing}},\ }\bibfield
  {title} {\bibinfo {title} {{Constraint on parameters of a rotating black hole
  in Einstein-bumblebee theory by quasi-periodic oscillations}},\ }\href
  {https://doi.org/10.1140/epjc/s10052-022-10475-x} {\bibfield  {journal}
  {\bibinfo  {journal} {Eur. Phys. J. C}\ }\textbf {\bibinfo {volume} {82}},\
  \bibinfo {pages} {528} (\bibinfo {year} {2022})},\ \Eprint
  {https://arxiv.org/abs/2112.02895} {arXiv:2112.02895 [gr-qc]} \BibitemShut
  {NoStop}%
\bibitem [{\citenamefont {Liu}\ \emph {et~al.}(2024{\natexlab{f}})\citenamefont
  {Liu}, \citenamefont {Fang}, \citenamefont {Jing},\ and\ \citenamefont
  {Wang}}]{Liu:2024oeq}%
  \BibitemOpen
  \bibfield  {author} {\bibinfo {author} {\bibfnamefont {W.}~\bibnamefont
  {Liu}}, \bibinfo {author} {\bibfnamefont {X.}~\bibnamefont {Fang}}, \bibinfo
  {author} {\bibfnamefont {J.}~\bibnamefont {Jing}},\ and\ \bibinfo {author}
  {\bibfnamefont {J.}~\bibnamefont {Wang}},\ }\bibfield  {title} {\bibinfo
  {title} {{Lorentz violation induces isospectrality breaking in
  Einstein-bumblebee gravity theory}},\ }\href
  {https://doi.org/10.1007/s11433-024-2405-y} {\bibfield  {journal} {\bibinfo
  {journal} {Sci. China Phys. Mech. Astron.}\ }\textbf {\bibinfo {volume}
  {67}},\ \bibinfo {pages} {280413} (\bibinfo {year} {2024}{\natexlab{f}})},\
  \Eprint {https://arxiv.org/abs/2402.09686} {arXiv:2402.09686 [gr-qc]}
  \BibitemShut {NoStop}%
\bibitem [{\citenamefont {Liu}\ \emph {et~al.}(2024{\natexlab{g}})\citenamefont
  {Liu}, \citenamefont {Wen},\ and\ \citenamefont {Wang}}]{Liu:2024wpa}%
  \BibitemOpen
  \bibfield  {author} {\bibinfo {author} {\bibfnamefont {W.}~\bibnamefont
  {Liu}}, \bibinfo {author} {\bibfnamefont {C.}~\bibnamefont {Wen}},\ and\
  \bibinfo {author} {\bibfnamefont {J.}~\bibnamefont {Wang}},\ }\bibfield
  {title} {\bibinfo {title} {{Lorentz violation alleviates gravitationally
  induced entanglement degradation}},\ }\href@noop {} {\  (\bibinfo {year}
  {2024}{\natexlab{g}})},\ \Eprint {https://arxiv.org/abs/2410.21681}
  {arXiv:2410.21681 [gr-qc]} \BibitemShut {NoStop}%
\bibitem [{\citenamefont {Mai}\ \emph {et~al.}(2024)\citenamefont {Mai},
  \citenamefont {Xu}, \citenamefont {Liang},\ and\ \citenamefont
  {Shao}}]{Mai:2024lgk}%
  \BibitemOpen
  \bibfield  {author} {\bibinfo {author} {\bibfnamefont {Z.-F.}\ \bibnamefont
  {Mai}}, \bibinfo {author} {\bibfnamefont {R.}~\bibnamefont {Xu}}, \bibinfo
  {author} {\bibfnamefont {D.}~\bibnamefont {Liang}},\ and\ \bibinfo {author}
  {\bibfnamefont {L.}~\bibnamefont {Shao}},\ }\bibfield  {title} {\bibinfo
  {title} {{Dynamic instability analysis for bumblebee black holes: The odd
  parity}},\ }\href {https://doi.org/10.1103/PhysRevD.109.084076} {\bibfield
  {journal} {\bibinfo  {journal} {Phys. Rev. D}\ }\textbf {\bibinfo {volume}
  {109}},\ \bibinfo {pages} {084076} (\bibinfo {year} {2024})},\ \Eprint
  {https://arxiv.org/abs/2401.07757} {arXiv:2401.07757 [gr-qc]} \BibitemShut
  {NoStop}%
\bibitem [{\citenamefont {Chen}\ \emph
  {et~al.}(2023{\natexlab{b}})\citenamefont {Chen}, \citenamefont {Pan},\ and\
  \citenamefont {Jing}}]{Chen:2023cjd}%
  \BibitemOpen
  \bibfield  {author} {\bibinfo {author} {\bibfnamefont {C.}~\bibnamefont
  {Chen}}, \bibinfo {author} {\bibfnamefont {Q.}~\bibnamefont {Pan}},\ and\
  \bibinfo {author} {\bibfnamefont {J.}~\bibnamefont {Jing}},\ }\bibfield
  {title} {\bibinfo {title} {{Quasinormal modes of a scalar perturbation around
  a rotating BTZ-like black hole in Einstein-bumblebee gravity}},\ }\href
  {https://doi.org/10.1016/j.physletb.2023.138186} {\bibfield  {journal}
  {\bibinfo  {journal} {Phys. Lett. B}\ }\textbf {\bibinfo {volume} {846}},\
  \bibinfo {pages} {138186} (\bibinfo {year} {2023}{\natexlab{b}})},\ \Eprint
  {https://arxiv.org/abs/2302.05861} {arXiv:2302.05861 [gr-qc]} \BibitemShut
  {NoStop}%
\bibitem [{\citenamefont {Ara\'ujo~Filho}(2025)}]{AraujoFilho:2025hkm}%
  \BibitemOpen
  \bibfield  {author} {\bibinfo {author} {\bibfnamefont {A.~A.}\ \bibnamefont
  {Ara\'ujo~Filho}},\ }\bibfield  {title} {\bibinfo {title} {{How does
  non-metricity affect particle creation and evaporation in bumblebee
  gravity?}},\ }\href@noop {} {\  (\bibinfo {year} {2025})},\ \Eprint
  {https://arxiv.org/abs/2501.00927} {arXiv:2501.00927 [gr-qc]} \BibitemShut
  {NoStop}%
\bibitem [{\citenamefont {Birrell}\ and\ \citenamefont
  {Davies}(1982)}]{birrell1984}%
  \BibitemOpen
  \bibfield  {author} {\bibinfo {author} {\bibfnamefont {N.~D.}\ \bibnamefont
  {Birrell}}\ and\ \bibinfo {author} {\bibfnamefont {P.~C.~W.}\ \bibnamefont
  {Davies}},\ }\href {https://doi.org/10.1017/CBO9780511622632} {\emph
  {\bibinfo {title} {{Quantum Fields in Curved Space}}}},\ Cambridge Monographs
  on Mathematical Physics\ (\bibinfo  {publisher} {Cambridge University
  Press},\ \bibinfo {address} {Cambridge, UK},\ \bibinfo {year}
  {1982})\BibitemShut {NoStop}%
\bibitem [{\citenamefont {Parker}\ and\ \citenamefont
  {Toms}(2009)}]{parker2009}%
  \BibitemOpen
  \bibfield  {author} {\bibinfo {author} {\bibfnamefont {L.~E.}\ \bibnamefont
  {Parker}}\ and\ \bibinfo {author} {\bibfnamefont {D.}~\bibnamefont {Toms}},\
  }\href {https://doi.org/10.1017/CBO9780511813924} {\emph {\bibinfo {title}
  {{Quantum Field Theory in Curved Spacetime}: {Quantized Field and
  Gravity}}}},\ Cambridge Monographs on Mathematical Physics\ (\bibinfo
  {publisher} {Cambridge University Press},\ \bibinfo {year}
  {2009})\BibitemShut {NoStop}%
\bibitem [{\citenamefont {Hawking}(1974)}]{hawking1974black}%
  \BibitemOpen
  \bibfield  {author} {\bibinfo {author} {\bibfnamefont {S.~W.}\ \bibnamefont
  {Hawking}},\ }\bibfield  {title} {\bibinfo {title} {{Black hole
  explosions}},\ }\href {https://doi.org/10.1038/248030a0} {\bibfield
  {journal} {\bibinfo  {journal} {Nature}\ }\textbf {\bibinfo {volume} {248}},\
  \bibinfo {pages} {30} (\bibinfo {year} {1974})}\BibitemShut {NoStop}%
\bibitem [{\citenamefont {Hawking}(1975)}]{hawking1975particle}%
  \BibitemOpen
  \bibfield  {author} {\bibinfo {author} {\bibfnamefont {S.~W.}\ \bibnamefont
  {Hawking}},\ }\bibfield  {title} {\bibinfo {title} {{Particle Creation by
  Black Holes}},\ }\href {https://doi.org/10.1007/BF02345020} {\bibfield
  {journal} {\bibinfo  {journal} {Commun. Math. Phys.}\ }\textbf {\bibinfo
  {volume} {43}},\ \bibinfo {pages} {199} (\bibinfo {year} {1975})},\ \bibinfo
  {note} {[Erratum: Commun.Math.Phys. 46, 206 (1976)]}\BibitemShut {NoStop}%
\bibitem [{\citenamefont {Hawking}(1976)}]{hawking1976black}%
  \BibitemOpen
  \bibfield  {author} {\bibinfo {author} {\bibfnamefont {S.~W.}\ \bibnamefont
  {Hawking}},\ }\bibfield  {title} {\bibinfo {title} {{Black Holes and
  Thermodynamics}},\ }\href {https://doi.org/10.1103/PhysRevD.13.191}
  {\bibfield  {journal} {\bibinfo  {journal} {Phys. Rev. D}\ }\textbf {\bibinfo
  {volume} {13}},\ \bibinfo {pages} {191} (\bibinfo {year} {1976})}\BibitemShut
  {NoStop}%
\bibitem [{\citenamefont {Wu}\ and\ \citenamefont {Zeng}(2022)}]{Wu:2022xwy}%
  \BibitemOpen
  \bibfield  {author} {\bibinfo {author} {\bibfnamefont {S.-M.}\ \bibnamefont
  {Wu}}\ and\ \bibinfo {author} {\bibfnamefont {H.-S.}\ \bibnamefont {Zeng}},\
  }\bibfield  {title} {\bibinfo {title} {{Genuine tripartite nonlocality and
  entanglement in curved spacetime}},\ }\href
  {https://doi.org/10.1140/epjc/s10052-021-09954-4} {\bibfield  {journal}
  {\bibinfo  {journal} {Eur. Phys. J. C}\ }\textbf {\bibinfo {volume} {82}},\
  \bibinfo {pages} {4} (\bibinfo {year} {2022})},\ \Eprint
  {https://arxiv.org/abs/2201.02333} {arXiv:2201.02333 [quant-ph]} \BibitemShut
  {NoStop}%
\bibitem [{\citenamefont {Wu}\ \emph {et~al.}(2023{\natexlab{a}})\citenamefont
  {Wu}, \citenamefont {Fan}, \citenamefont {Wang}, \citenamefont {Wu},
  \citenamefont {Huang},\ and\ \citenamefont {Zeng}}]{Wu:2023sye}%
  \BibitemOpen
  \bibfield  {author} {\bibinfo {author} {\bibfnamefont {S.-M.}\ \bibnamefont
  {Wu}}, \bibinfo {author} {\bibfnamefont {X.-W.}\ \bibnamefont {Fan}},
  \bibinfo {author} {\bibfnamefont {R.-D.}\ \bibnamefont {Wang}}, \bibinfo
  {author} {\bibfnamefont {H.-Y.}\ \bibnamefont {Wu}}, \bibinfo {author}
  {\bibfnamefont {X.-L.}\ \bibnamefont {Huang}},\ and\ \bibinfo {author}
  {\bibfnamefont {H.-S.}\ \bibnamefont {Zeng}},\ }\bibfield  {title} {\bibinfo
  {title} {{Does Hawking effect always degrade fidelity of quantum
  teleportation in Schwarzschild spacetime?}},\ }\href
  {https://doi.org/10.1007/JHEP11(2023)232} {\bibfield  {journal} {\bibinfo
  {journal} {JHEP}\ }\textbf {\bibinfo {volume} {11}},\ \bibinfo {pages}
  {232}},\ \Eprint {https://arxiv.org/abs/2304.00984} {arXiv:2304.00984
  [gr-qc]} \BibitemShut {NoStop}%
\bibitem [{\citenamefont {Liu}\ \emph {et~al.}(2024{\natexlab{h}})\citenamefont
  {Liu}, \citenamefont {Zeng},\ and\ \citenamefont {Wang}}]{Liu:2024yrf}%
  \BibitemOpen
  \bibfield  {author} {\bibinfo {author} {\bibfnamefont {X.}~\bibnamefont
  {Liu}}, \bibinfo {author} {\bibfnamefont {C.}~\bibnamefont {Zeng}},\ and\
  \bibinfo {author} {\bibfnamefont {J.}~\bibnamefont {Wang}},\ }\bibfield
  {title} {\bibinfo {title} {{Generation of quantum entanglement in superposed
  diamond spacetime}},\ }\href@noop {} {\  (\bibinfo {year}
  {2024}{\natexlab{h}})},\ \Eprint {https://arxiv.org/abs/2501.00246}
  {arXiv:2501.00246 [gr-qc]} \BibitemShut {NoStop}%
\bibitem [{\citenamefont {Unruh}(1976)}]{unruh1976notes}%
  \BibitemOpen
  \bibfield  {author} {\bibinfo {author} {\bibfnamefont {W.~G.}\ \bibnamefont
  {Unruh}},\ }\bibfield  {title} {\bibinfo {title} {{Notes on black hole
  evaporation}},\ }\href {https://doi.org/10.1103/PhysRevD.14.870} {\bibfield
  {journal} {\bibinfo  {journal} {Phys. Rev. D}\ }\textbf {\bibinfo {volume}
  {14}},\ \bibinfo {pages} {870} (\bibinfo {year} {1976})}\BibitemShut
  {NoStop}%
\bibitem [{\citenamefont {Crispino}\ \emph {et~al.}(2008)\citenamefont
  {Crispino}, \citenamefont {Higuchi},\ and\ \citenamefont
  {Matsas}}]{crispino2008unruh}%
  \BibitemOpen
  \bibfield  {author} {\bibinfo {author} {\bibfnamefont {L.~C.~B.}\
  \bibnamefont {Crispino}}, \bibinfo {author} {\bibfnamefont {A.}~\bibnamefont
  {Higuchi}},\ and\ \bibinfo {author} {\bibfnamefont {G.~E.~A.}\ \bibnamefont
  {Matsas}},\ }\bibfield  {title} {\bibinfo {title} {{The Unruh effect and its
  applications}},\ }\href {https://doi.org/10.1103/RevModPhys.80.787}
  {\bibfield  {journal} {\bibinfo  {journal} {Rev. Mod. Phys.}\ }\textbf
  {\bibinfo {volume} {80}},\ \bibinfo {pages} {787} (\bibinfo {year} {2008})},\
  \Eprint {https://arxiv.org/abs/0710.5373} {arXiv:0710.5373 [gr-qc]}
  \BibitemShut {NoStop}%
\bibitem [{\citenamefont {Wu}\ \emph {et~al.}(2023{\natexlab{b}})\citenamefont
  {Wu}, \citenamefont {Wang}, \citenamefont {Liu}, \citenamefont {Huang},\ and\
  \citenamefont {Zeng}}]{Wu:2022lmc}%
  \BibitemOpen
  \bibfield  {author} {\bibinfo {author} {\bibfnamefont {S.-M.}\ \bibnamefont
  {Wu}}, \bibinfo {author} {\bibfnamefont {C.-X.}\ \bibnamefont {Wang}},
  \bibinfo {author} {\bibfnamefont {D.-D.}\ \bibnamefont {Liu}}, \bibinfo
  {author} {\bibfnamefont {X.-L.}\ \bibnamefont {Huang}},\ and\ \bibinfo
  {author} {\bibfnamefont {H.-S.}\ \bibnamefont {Zeng}},\ }\bibfield  {title}
  {\bibinfo {title} {{Would quantum coherence be increased by curvature effect
  in de Sitter space?}},\ }\href {https://doi.org/10.1007/JHEP02(2023)115}
  {\bibfield  {journal} {\bibinfo  {journal} {JHEP}\ }\textbf {\bibinfo
  {volume} {02}},\ \bibinfo {pages} {115}},\ \Eprint
  {https://arxiv.org/abs/2207.11721} {arXiv:2207.11721 [gr-qc]} \BibitemShut
  {NoStop}%
\bibitem [{\citenamefont {Wu}\ \emph {et~al.}(2024{\natexlab{a}})\citenamefont
  {Wu}, \citenamefont {Teng}, \citenamefont {Li}, \citenamefont {Li},
  \citenamefont {Liu},\ and\ \citenamefont {Wang}}]{Wu:2023spa}%
  \BibitemOpen
  \bibfield  {author} {\bibinfo {author} {\bibfnamefont {S.-M.}\ \bibnamefont
  {Wu}}, \bibinfo {author} {\bibfnamefont {X.-W.}\ \bibnamefont {Teng}},
  \bibinfo {author} {\bibfnamefont {J.-X.}\ \bibnamefont {Li}}, \bibinfo
  {author} {\bibfnamefont {S.-H.}\ \bibnamefont {Li}}, \bibinfo {author}
  {\bibfnamefont {T.-H.}\ \bibnamefont {Liu}},\ and\ \bibinfo {author}
  {\bibfnamefont {J.-C.}\ \bibnamefont {Wang}},\ }\bibfield  {title} {\bibinfo
  {title} {{Genuinely accessible and inaccessible entanglement in Schwarzschild
  black hole}},\ }\href {https://doi.org/10.1016/j.physletb.2023.138334}
  {\bibfield  {journal} {\bibinfo  {journal} {Phys. Lett. B}\ }\textbf
  {\bibinfo {volume} {848}},\ \bibinfo {pages} {138334} (\bibinfo {year}
  {2024}{\natexlab{a}})},\ \Eprint {https://arxiv.org/abs/2311.12362}
  {arXiv:2311.12362 [gr-qc]} \BibitemShut {NoStop}%
\bibitem [{\citenamefont {Li}\ \emph {et~al.}(2024)\citenamefont {Li},
  \citenamefont {Wu}, \citenamefont {Zeng},\ and\ \citenamefont
  {Huang}}]{Li:2022pwa}%
  \BibitemOpen
  \bibfield  {author} {\bibinfo {author} {\bibfnamefont {W.-M.}\ \bibnamefont
  {Li}}, \bibinfo {author} {\bibfnamefont {S.-M.}\ \bibnamefont {Wu}}, \bibinfo
  {author} {\bibfnamefont {H.-S.}\ \bibnamefont {Zeng}},\ and\ \bibinfo
  {author} {\bibfnamefont {X.-L.}\ \bibnamefont {Huang}},\ }\bibfield  {title}
  {\bibinfo {title} {{Bosonic and fermionic coherence of N-partite states in
  the background of a dilaton black hole}},\ }\href
  {https://doi.org/10.1007/JHEP09(2024)144} {\bibfield  {journal} {\bibinfo
  {journal} {JHEP}\ }\textbf {\bibinfo {volume} {09}},\ \bibinfo {pages}
  {144}},\ \Eprint {https://arxiv.org/abs/2407.07688} {arXiv:2407.07688
  [gr-qc]} \BibitemShut {NoStop}%
\bibitem [{\citenamefont {Moore}(1970)}]{moore:1970tmc}%
  \BibitemOpen
  \bibfield  {author} {\bibinfo {author} {\bibfnamefont {G.~T.}\ \bibnamefont
  {Moore}},\ }\bibfield  {title} {\bibinfo {title} {{Quantum Theory of the
  Electromagnetic Field in a Variable-Length One-Dimensional Cavity}},\ }\href
  {https://doi.org/10.1063/1.1665432} {\bibfield  {journal} {\bibinfo
  {journal} {J. Math. Phys.}\ }\textbf {\bibinfo {volume} {11}},\ \bibinfo
  {pages} {2679} (\bibinfo {year} {1970})}\BibitemShut {NoStop}%
\bibitem [{\citenamefont {Anderson}\ \emph {et~al.}(2017)\citenamefont
  {Anderson}, \citenamefont {Good},\ and\ \citenamefont
  {Evans}}]{anderson:2015iga}%
  \BibitemOpen
  \bibfield  {author} {\bibinfo {author} {\bibfnamefont {P.~R.}\ \bibnamefont
  {Anderson}}, \bibinfo {author} {\bibfnamefont {M.~R.~R.}\ \bibnamefont
  {Good}},\ and\ \bibinfo {author} {\bibfnamefont {C.~R.}\ \bibnamefont
  {Evans}},\ }\bibfield  {title} {\bibinfo {title} {{Black hole - moving mirror
  I: An exact correspondence}},\ }in\ \href
  {https://doi.org/10.1142/9789813226609_0171} {\emph {\bibinfo {booktitle}
  {{14th Marcel Grossmann Meeting on Recent Developments in Theoretical and
  Experimental General Relativity, Astrophysics, and Relativistic Field
  Theories}}}},\ Vol.~\bibinfo {volume} {2}\ (\bibinfo {year} {2017})\ pp.\
  \bibinfo {pages} {1701--1704},\ \Eprint {https://arxiv.org/abs/1507.03489}
  {arXiv:1507.03489 [gr-qc]} \BibitemShut {NoStop}%
\bibitem [{\citenamefont {Good}\ \emph {et~al.}(2017)\citenamefont {Good},
  \citenamefont {Anderson},\ and\ \citenamefont {Evans}}]{good:2015jwa}%
  \BibitemOpen
  \bibfield  {author} {\bibinfo {author} {\bibfnamefont {M.~R.~R.}\
  \bibnamefont {Good}}, \bibinfo {author} {\bibfnamefont {P.~R.}\ \bibnamefont
  {Anderson}},\ and\ \bibinfo {author} {\bibfnamefont {C.~R.}\ \bibnamefont
  {Evans}},\ }\bibfield  {title} {\bibinfo {title} {{Black hole - moving mirror
  II: Particle creation}},\ }in\ \href
  {https://doi.org/10.1142/9789813226609_0172} {\emph {\bibinfo {booktitle}
  {{14th Marcel Grossmann Meeting on Recent Developments in Theoretical and
  Experimental General Relativity, Astrophysics, and Relativistic Field
  Theories}}}},\ Vol.~\bibinfo {volume} {2}\ (\bibinfo {year} {2017})\ pp.\
  \bibinfo {pages} {1705--1708},\ \Eprint {https://arxiv.org/abs/1507.05048}
  {arXiv:1507.05048 [gr-qc]} \BibitemShut {NoStop}%
\bibitem [{\citenamefont {Scully}\ \emph {et~al.}(2003)\citenamefont {Scully},
  \citenamefont {Kocharovsky}, \citenamefont {Belyanin}, \citenamefont {Fry},\
  and\ \citenamefont {Capasso}}]{Scully:2003zz}%
  \BibitemOpen
  \bibfield  {author} {\bibinfo {author} {\bibfnamefont {M.~O.}\ \bibnamefont
  {Scully}}, \bibinfo {author} {\bibfnamefont {V.~V.}\ \bibnamefont
  {Kocharovsky}}, \bibinfo {author} {\bibfnamefont {A.}~\bibnamefont
  {Belyanin}}, \bibinfo {author} {\bibfnamefont {E.}~\bibnamefont {Fry}},\ and\
  \bibinfo {author} {\bibfnamefont {F.}~\bibnamefont {Capasso}},\ }\bibfield
  {title} {\bibinfo {title} {{Enhancing Acceleration Radiation from
  Ground-State Atoms via Cavity Quantum Electrodynamics}},\ }\href
  {https://doi.org/10.1103/PhysRevLett.91.243004} {\bibfield  {journal}
  {\bibinfo  {journal} {Phys. Rev. Lett.}\ }\textbf {\bibinfo {volume} {91}},\
  \bibinfo {pages} {243004} (\bibinfo {year} {2003})},\ \Eprint
  {https://arxiv.org/abs/quant-ph/0305178} {arXiv:quant-ph/0305178}
  \BibitemShut {NoStop}%
\bibitem [{\citenamefont {Scully}\ \emph {et~al.}(2018)\citenamefont {Scully},
  \citenamefont {Fulling}, \citenamefont {Lee}, \citenamefont {Page},
  \citenamefont {Schleich},\ and\ \citenamefont
  {Svidzinsky}}]{scully2018quantum}%
  \BibitemOpen
  \bibfield  {author} {\bibinfo {author} {\bibfnamefont {M.~O.}\ \bibnamefont
  {Scully}}, \bibinfo {author} {\bibfnamefont {S.}~\bibnamefont {Fulling}},
  \bibinfo {author} {\bibfnamefont {D.}~\bibnamefont {Lee}}, \bibinfo {author}
  {\bibfnamefont {D.~N.}\ \bibnamefont {Page}}, \bibinfo {author}
  {\bibfnamefont {W.}~\bibnamefont {Schleich}},\ and\ \bibinfo {author}
  {\bibfnamefont {A.}~\bibnamefont {Svidzinsky}},\ }\bibfield  {title}
  {\bibinfo {title} {{Quantum optics approach to radiation from atoms falling
  into a black hole}},\ }\href {https://doi.org/10.1073/pnas.1807703115}
  {\bibfield  {journal} {\bibinfo  {journal} {Proc. Nat. Acad. Sci.}\ }\textbf
  {\bibinfo {volume} {115}},\ \bibinfo {pages} {8131} (\bibinfo {year}
  {2018})},\ \Eprint {https://arxiv.org/abs/1709.00481} {arXiv:1709.00481
  [quant-ph]} \BibitemShut {NoStop}%
\bibitem [{\citenamefont {Sen}\ \emph {et~al.}(2022{\natexlab{a}})\citenamefont
  {Sen}, \citenamefont {Mandal},\ and\ \citenamefont
  {Gangopadhyay}}]{Sen:2022cdx}%
  \BibitemOpen
  \bibfield  {author} {\bibinfo {author} {\bibfnamefont {S.}~\bibnamefont
  {Sen}}, \bibinfo {author} {\bibfnamefont {R.}~\bibnamefont {Mandal}},\ and\
  \bibinfo {author} {\bibfnamefont {S.}~\bibnamefont {Gangopadhyay}},\
  }\bibfield  {title} {\bibinfo {title} {{Near horizon aspects of acceleration
  radiation of an atom falling into a class of static spherically symmetric
  black hole geometries}},\ }\href
  {https://doi.org/10.1103/PhysRevD.106.025004} {\bibfield  {journal} {\bibinfo
   {journal} {Phys. Rev. D}\ }\textbf {\bibinfo {volume} {106}},\ \bibinfo
  {pages} {025004} (\bibinfo {year} {2022}{\natexlab{a}})},\ \Eprint
  {https://arxiv.org/abs/2205.11260} {arXiv:2205.11260 [gr-qc]} \BibitemShut
  {NoStop}%
\bibitem [{\citenamefont {Azizi}\ \emph
  {et~al.}(2021{\natexlab{a}})\citenamefont {Azizi}, \citenamefont {Camblong},
  \citenamefont {Chakraborty}, \citenamefont {Ordonez},\ and\ \citenamefont
  {Scully}}]{Azizi:2020gff}%
  \BibitemOpen
  \bibfield  {author} {\bibinfo {author} {\bibfnamefont {A.}~\bibnamefont
  {Azizi}}, \bibinfo {author} {\bibfnamefont {H.~E.}\ \bibnamefont {Camblong}},
  \bibinfo {author} {\bibfnamefont {A.}~\bibnamefont {Chakraborty}}, \bibinfo
  {author} {\bibfnamefont {C.~R.}\ \bibnamefont {Ordonez}},\ and\ \bibinfo
  {author} {\bibfnamefont {M.~O.}\ \bibnamefont {Scully}},\ }\bibfield  {title}
  {\bibinfo {title} {{Acceleration radiation of an atom freely falling into a
  Kerr black hole and near-horizon conformal quantum mechanics}},\ }\href
  {https://doi.org/10.1103/PhysRevD.104.065006} {\bibfield  {journal} {\bibinfo
   {journal} {Phys. Rev. D}\ }\textbf {\bibinfo {volume} {104}},\ \bibinfo
  {pages} {065006} (\bibinfo {year} {2021}{\natexlab{a}})},\ \Eprint
  {https://arxiv.org/abs/2011.08368} {arXiv:2011.08368 [gr-qc]} \BibitemShut
  {NoStop}%
\bibitem [{\citenamefont {Sen}\ \emph {et~al.}(2023)\citenamefont {Sen},
  \citenamefont {Mandal},\ and\ \citenamefont {Gangopadhyay}}]{Sen:2023zfq}%
  \BibitemOpen
  \bibfield  {author} {\bibinfo {author} {\bibfnamefont {S.}~\bibnamefont
  {Sen}}, \bibinfo {author} {\bibfnamefont {R.}~\bibnamefont {Mandal}},\ and\
  \bibinfo {author} {\bibfnamefont {S.}~\bibnamefont {Gangopadhyay}},\
  }\bibfield  {title} {\bibinfo {title} {{Near horizon approximation and beyond
  for a two-level atom falling into a Kerr\textendash{}Newman black hole}},\
  }\href {https://doi.org/10.1140/epjp/s13360-023-04482-4} {\bibfield
  {journal} {\bibinfo  {journal} {Eur. Phys. J. Plus}\ }\textbf {\bibinfo
  {volume} {138}},\ \bibinfo {pages} {855} (\bibinfo {year} {2023})},\ \Eprint
  {https://arxiv.org/abs/2301.04834} {arXiv:2301.04834 [gr-qc]} \BibitemShut
  {NoStop}%
\bibitem [{\citenamefont {Sen}\ \emph {et~al.}(2022{\natexlab{b}})\citenamefont
  {Sen}, \citenamefont {Mandal},\ and\ \citenamefont
  {Gangopadhyay}}]{Sen:2022tru}%
  \BibitemOpen
  \bibfield  {author} {\bibinfo {author} {\bibfnamefont {S.}~\bibnamefont
  {Sen}}, \bibinfo {author} {\bibfnamefont {R.}~\bibnamefont {Mandal}},\ and\
  \bibinfo {author} {\bibfnamefont {S.}~\bibnamefont {Gangopadhyay}},\
  }\bibfield  {title} {\bibinfo {title} {{Equivalence principle and HBAR
  entropy of an atom falling into a quantum corrected black hole}},\ }\href
  {https://doi.org/10.1103/PhysRevD.105.085007} {\bibfield  {journal} {\bibinfo
   {journal} {Phys. Rev. D}\ }\textbf {\bibinfo {volume} {105}},\ \bibinfo
  {pages} {085007} (\bibinfo {year} {2022}{\natexlab{b}})},\ \Eprint
  {https://arxiv.org/abs/2202.00671} {arXiv:2202.00671 [hep-th]} \BibitemShut
  {NoStop}%
\bibitem [{\citenamefont {Bukhari}\ \emph {et~al.}(2023)\citenamefont
  {Bukhari}, \citenamefont {Bhat}, \citenamefont {Xu},\ and\ \citenamefont
  {Wang}}]{Bukhari:2022wyx}%
  \BibitemOpen
  \bibfield  {author} {\bibinfo {author} {\bibfnamefont {S.~M. A.~S.}\
  \bibnamefont {Bukhari}}, \bibinfo {author} {\bibfnamefont {I.~A.}\
  \bibnamefont {Bhat}}, \bibinfo {author} {\bibfnamefont {C.}~\bibnamefont
  {Xu}},\ and\ \bibinfo {author} {\bibfnamefont {L.-G.}\ \bibnamefont {Wang}},\
  }\bibfield  {title} {\bibinfo {title} {{Nonthermal acceleration radiation of
  atoms near a black hole in presence of dark energy}},\ }\href
  {https://doi.org/10.1103/PhysRevD.107.105017} {\bibfield  {journal} {\bibinfo
   {journal} {Phys. Rev. D}\ }\textbf {\bibinfo {volume} {107}},\ \bibinfo
  {pages} {105017} (\bibinfo {year} {2023})},\ \Eprint
  {https://arxiv.org/abs/2211.08793} {arXiv:2211.08793 [gr-qc]} \BibitemShut
  {NoStop}%
\bibitem [{\citenamefont {Bukhari}\ and\ \citenamefont
  {Wang}(2024)}]{Bukhari:2023yuy}%
  \BibitemOpen
  \bibfield  {author} {\bibinfo {author} {\bibfnamefont {S.~M. A.~S.}\
  \bibnamefont {Bukhari}}\ and\ \bibinfo {author} {\bibfnamefont {L.-G.}\
  \bibnamefont {Wang}},\ }\bibfield  {title} {\bibinfo {title} {{Seeing dark
  matter via acceleration radiation}},\ }\href
  {https://doi.org/10.1103/PhysRevD.109.045009} {\bibfield  {journal} {\bibinfo
   {journal} {Phys. Rev. D}\ }\textbf {\bibinfo {volume} {109}},\ \bibinfo
  {pages} {045009} (\bibinfo {year} {2024})},\ \Eprint
  {https://arxiv.org/abs/2309.11958} {arXiv:2309.11958 [gr-qc]} \BibitemShut
  {NoStop}%
\bibitem [{\citenamefont {Jana}\ \emph {et~al.}(2024)\citenamefont {Jana},
  \citenamefont {Sen},\ and\ \citenamefont {Gangopadhyay}}]{Jana:2024fhx}%
  \BibitemOpen
  \bibfield  {author} {\bibinfo {author} {\bibfnamefont {A.}~\bibnamefont
  {Jana}}, \bibinfo {author} {\bibfnamefont {S.}~\bibnamefont {Sen}},\ and\
  \bibinfo {author} {\bibfnamefont {S.}~\bibnamefont {Gangopadhyay}},\
  }\bibfield  {title} {\bibinfo {title} {{Atom falling into a quantum corrected
  charged black hole and HBAR entropy}},\ }\href
  {https://doi.org/10.1103/PhysRevD.110.026029} {\bibfield  {journal} {\bibinfo
   {journal} {Phys. Rev. D}\ }\textbf {\bibinfo {volume} {110}},\ \bibinfo
  {pages} {026029} (\bibinfo {year} {2024})},\ \Eprint
  {https://arxiv.org/abs/2405.13087} {arXiv:2405.13087 [gr-qc]} \BibitemShut
  {NoStop}%
\bibitem [{\citenamefont {Das}\ \emph {et~al.}(2024)\citenamefont {Das},
  \citenamefont {Sen},\ and\ \citenamefont {Gangopadhyay}}]{Das:2023rwg}%
  \BibitemOpen
  \bibfield  {author} {\bibinfo {author} {\bibfnamefont {A.}~\bibnamefont
  {Das}}, \bibinfo {author} {\bibfnamefont {S.}~\bibnamefont {Sen}},\ and\
  \bibinfo {author} {\bibfnamefont {S.}~\bibnamefont {Gangopadhyay}},\
  }\bibfield  {title} {\bibinfo {title} {{Horizon brightened accelerated
  radiation in the background of braneworld black holes}},\ }\href
  {https://doi.org/10.1103/PhysRevD.109.064087} {\bibfield  {journal} {\bibinfo
   {journal} {Phys. Rev. D}\ }\textbf {\bibinfo {volume} {109}},\ \bibinfo
  {pages} {064087} (\bibinfo {year} {2024})},\ \Eprint
  {https://arxiv.org/abs/2311.13557} {arXiv:2311.13557 [gr-qc]} \BibitemShut
  {NoStop}%
\bibitem [{\citenamefont {Masood}\ and\ \citenamefont
  {Mikki}(2024)}]{Masood:2024glj}%
  \BibitemOpen
  \bibfield  {author} {\bibinfo {author} {\bibfnamefont {S.}~\bibnamefont
  {Masood}}\ and\ \bibinfo {author} {\bibfnamefont {S.}~\bibnamefont {Mikki}},\
  }\bibfield  {title} {\bibinfo {title} {{A Casimir-like probe for 4D
  Einstein-Gauss-Bonnet gravity}},\ }\href@noop {} {\  (\bibinfo {year}
  {2024})},\ \Eprint {https://arxiv.org/abs/2407.02313} {arXiv:2407.02313
  [gr-qc]} \BibitemShut {NoStop}%
\bibitem [{\citenamefont
  {Chandrasekhar}(1985)}]{chandrasekhar1998mathematical}%
  \BibitemOpen
  \bibfield  {author} {\bibinfo {author} {\bibfnamefont {S.}~\bibnamefont
  {Chandrasekhar}},\ }\href@noop {} {\emph {\bibinfo {title} {{The mathematical
  theory of black holes}}}}\ (\bibinfo {year} {1985})\BibitemShut {NoStop}%
\bibitem [{\citenamefont {Camblong}\ and\ \citenamefont
  {Ordonez}(2005{\natexlab{a}})}]{camblong2005black}%
  \BibitemOpen
  \bibfield  {author} {\bibinfo {author} {\bibfnamefont {H.~E.}\ \bibnamefont
  {Camblong}}\ and\ \bibinfo {author} {\bibfnamefont {C.~R.}\ \bibnamefont
  {Ordonez}},\ }\bibfield  {title} {\bibinfo {title} {{Black hole
  thermodynamics from near-horizon conformal quantum mechanics}},\ }\href
  {https://doi.org/10.1103/PhysRevD.71.104029} {\bibfield  {journal} {\bibinfo
  {journal} {Phys. Rev. D}\ }\textbf {\bibinfo {volume} {71}},\ \bibinfo
  {pages} {104029} (\bibinfo {year} {2005}{\natexlab{a}})},\ \Eprint
  {https://arxiv.org/abs/hep-th/0411008} {arXiv:hep-th/0411008} \BibitemShut
  {NoStop}%
\bibitem [{\citenamefont {Camblong}\ and\ \citenamefont
  {Ordonez}(2005{\natexlab{b}})}]{camblong2005semiclassical}%
  \BibitemOpen
  \bibfield  {author} {\bibinfo {author} {\bibfnamefont {H.~E.}\ \bibnamefont
  {Camblong}}\ and\ \bibinfo {author} {\bibfnamefont {C.~R.}\ \bibnamefont
  {Ordonez}},\ }\bibfield  {title} {\bibinfo {title} {{Semiclassical methods in
  curved spacetime and black hole thermodynamics}},\ }\href
  {https://doi.org/10.1103/PhysRevD.71.124040} {\bibfield  {journal} {\bibinfo
  {journal} {Phys. Rev. D}\ }\textbf {\bibinfo {volume} {71}},\ \bibinfo
  {pages} {124040} (\bibinfo {year} {2005}{\natexlab{b}})},\ \Eprint
  {https://arxiv.org/abs/hep-th/0412309} {arXiv:hep-th/0412309} \BibitemShut
  {NoStop}%
\bibitem [{\citenamefont {Camblong}\ \emph {et~al.}(2020)\citenamefont
  {Camblong}, \citenamefont {Chakraborty},\ and\ \citenamefont
  {Ordonez}}]{camblong2020near}%
  \BibitemOpen
  \bibfield  {author} {\bibinfo {author} {\bibfnamefont {H.~E.}\ \bibnamefont
  {Camblong}}, \bibinfo {author} {\bibfnamefont {A.}~\bibnamefont
  {Chakraborty}},\ and\ \bibinfo {author} {\bibfnamefont {C.~R.}\ \bibnamefont
  {Ordonez}},\ }\bibfield  {title} {\bibinfo {title} {{Near-horizon aspects of
  acceleration radiation by free fall of an atom into a black hole}},\ }\href
  {https://doi.org/10.1103/PhysRevD.102.085010} {\bibfield  {journal} {\bibinfo
   {journal} {Phys. Rev. D}\ }\textbf {\bibinfo {volume} {102}},\ \bibinfo
  {pages} {085010} (\bibinfo {year} {2020})},\ \Eprint
  {https://arxiv.org/abs/2009.06580} {arXiv:2009.06580 [gr-qc]} \BibitemShut
  {NoStop}%
\bibitem [{\citenamefont {Boulware}(1975)}]{boulware1975quantum}%
  \BibitemOpen
  \bibfield  {author} {\bibinfo {author} {\bibfnamefont {D.~G.}\ \bibnamefont
  {Boulware}},\ }\bibfield  {title} {\bibinfo {title} {{Quantum Field Theory in
  Schwarzschild and Rindler Spaces}},\ }\href
  {https://doi.org/10.1103/PhysRevD.11.1404} {\bibfield  {journal} {\bibinfo
  {journal} {Phys. Rev. D}\ }\textbf {\bibinfo {volume} {11}},\ \bibinfo
  {pages} {1404} (\bibinfo {year} {1975})}\BibitemShut {NoStop}%
\bibitem [{\citenamefont {Camblong}\ and\ \citenamefont
  {Ordonez}(2003)}]{camblong2003anomaly}%
  \BibitemOpen
  \bibfield  {author} {\bibinfo {author} {\bibfnamefont {H.~E.}\ \bibnamefont
  {Camblong}}\ and\ \bibinfo {author} {\bibfnamefont {C.~R.}\ \bibnamefont
  {Ordonez}},\ }\bibfield  {title} {\bibinfo {title} {{Anomaly in conformal
  quantum mechanics: From molecular physics to black holes}},\ }\href
  {https://doi.org/10.1103/PhysRevD.68.125013} {\bibfield  {journal} {\bibinfo
  {journal} {Phys. Rev. D}\ }\textbf {\bibinfo {volume} {68}},\ \bibinfo
  {pages} {125013} (\bibinfo {year} {2003})},\ \Eprint
  {https://arxiv.org/abs/hep-th/0303166} {arXiv:hep-th/0303166} \BibitemShut
  {NoStop}%
\bibitem [{\citenamefont {Azizi}\ \emph
  {et~al.}(2021{\natexlab{b}})\citenamefont {Azizi}, \citenamefont {Camblong},
  \citenamefont {Chakraborty}, \citenamefont {Ordonez},\ and\ \citenamefont
  {Scully}}]{azizi2021quantum}%
  \BibitemOpen
  \bibfield  {author} {\bibinfo {author} {\bibfnamefont {A.}~\bibnamefont
  {Azizi}}, \bibinfo {author} {\bibfnamefont {H.~E.}\ \bibnamefont {Camblong}},
  \bibinfo {author} {\bibfnamefont {A.}~\bibnamefont {Chakraborty}}, \bibinfo
  {author} {\bibfnamefont {C.~R.}\ \bibnamefont {Ordonez}},\ and\ \bibinfo
  {author} {\bibfnamefont {M.~O.}\ \bibnamefont {Scully}},\ }\bibfield  {title}
  {\bibinfo {title} {{Quantum optics meets black hole thermodynamics via
  conformal quantum mechanics: I. Master equation for acceleration
  radiation}},\ }\bibfield  {journal} {\bibinfo  {journal} {Phys. Rev. D}\
  }\textbf {\bibinfo {volume} {104}},\ \href
  {https://doi.org/10.1103/PhysRevD.104.084086} {10.1103/PhysRevD.104.084086}
  (\bibinfo {year} {2021}{\natexlab{b}}),\ \Eprint
  {https://arxiv.org/abs/2108.07570} {arXiv:2108.07570 [gr-qc]} \BibitemShut
  {NoStop}%
\bibitem [{\citenamefont {Chakraborty}\ and\ \citenamefont
  {Majhi}(2019)}]{chakraborty2019detector}%
  \BibitemOpen
  \bibfield  {author} {\bibinfo {author} {\bibfnamefont {K.}~\bibnamefont
  {Chakraborty}}\ and\ \bibinfo {author} {\bibfnamefont {B.~R.}\ \bibnamefont
  {Majhi}},\ }\bibfield  {title} {\bibinfo {title} {{Detector response along
  null geodesics in black hole spacetimes and in a
  Friedmann-Lemaitre-Robertson-Walker Universe}},\ }\href
  {https://doi.org/10.1103/PhysRevD.100.045004} {\bibfield  {journal} {\bibinfo
   {journal} {Phys. Rev. D}\ }\textbf {\bibinfo {volume} {100}},\ \bibinfo
  {pages} {045004} (\bibinfo {year} {2019})},\ \Eprint
  {https://arxiv.org/abs/1905.10554} {arXiv:1905.10554 [gr-qc]} \BibitemShut
  {NoStop}%
\bibitem [{\citenamefont {Wei}\ \emph {et~al.}(2022)\citenamefont {Wei},
  \citenamefont {Liu},\ and\ \citenamefont {Mann}}]{Wei:2022dzw}%
  \BibitemOpen
  \bibfield  {author} {\bibinfo {author} {\bibfnamefont {S.-W.}\ \bibnamefont
  {Wei}}, \bibinfo {author} {\bibfnamefont {Y.-X.}\ \bibnamefont {Liu}},\ and\
  \bibinfo {author} {\bibfnamefont {R.~B.}\ \bibnamefont {Mann}},\ }\bibfield
  {title} {\bibinfo {title} {{Black Hole Solutions as Topological Thermodynamic
  Defects}},\ }\href {https://doi.org/10.1103/PhysRevLett.129.191101}
  {\bibfield  {journal} {\bibinfo  {journal} {Phys. Rev. Lett.}\ }\textbf
  {\bibinfo {volume} {129}},\ \bibinfo {pages} {191101} (\bibinfo {year}
  {2022})},\ \Eprint {https://arxiv.org/abs/2208.01932} {arXiv:2208.01932
  [gr-qc]} \BibitemShut {NoStop}%
\bibitem [{\citenamefont {Wu}(2023{\natexlab{a}})}]{Wu:2022whe}%
  \BibitemOpen
  \bibfield  {author} {\bibinfo {author} {\bibfnamefont {D.}~\bibnamefont
  {Wu}},\ }\bibfield  {title} {\bibinfo {title} {{Topological classes of
  rotating black holes}},\ }\href {https://doi.org/10.1103/PhysRevD.107.024024}
  {\bibfield  {journal} {\bibinfo  {journal} {Phys. Rev. D}\ }\textbf {\bibinfo
  {volume} {107}},\ \bibinfo {pages} {024024} (\bibinfo {year}
  {2023}{\natexlab{a}})},\ \Eprint {https://arxiv.org/abs/2211.15151}
  {arXiv:2211.15151 [gr-qc]} \BibitemShut {NoStop}%
\bibitem [{\citenamefont {Wu}(2023{\natexlab{b}})}]{Wu:2023xpq}%
  \BibitemOpen
  \bibfield  {author} {\bibinfo {author} {\bibfnamefont {D.}~\bibnamefont
  {Wu}},\ }\bibfield  {title} {\bibinfo {title} {{Classifying topology of
  consistent thermodynamics of the four-dimensional neutral Lorentzian
  NUT-charged spacetimes}},\ }\href
  {https://doi.org/10.1140/epjc/s10052-023-11561-4} {\bibfield  {journal}
  {\bibinfo  {journal} {Eur. Phys. J. C}\ }\textbf {\bibinfo {volume} {83}},\
  \bibinfo {pages} {365} (\bibinfo {year} {2023}{\natexlab{b}})},\ \Eprint
  {https://arxiv.org/abs/2302.01100} {arXiv:2302.01100 [gr-qc]} \BibitemShut
  {NoStop}%
\bibitem [{\citenamefont {Wu}(2023{\natexlab{c}})}]{Wu:2023fcw}%
  \BibitemOpen
  \bibfield  {author} {\bibinfo {author} {\bibfnamefont {D.}~\bibnamefont
  {Wu}},\ }\bibfield  {title} {\bibinfo {title} {{Consistent thermodynamics and
  topological classes for the four-dimensional Lorentzian charged Taub-NUT
  spacetimes}},\ }\href {https://doi.org/10.1140/epjc/s10052-023-11782-7}
  {\bibfield  {journal} {\bibinfo  {journal} {Eur. Phys. J. C}\ }\textbf
  {\bibinfo {volume} {83}},\ \bibinfo {pages} {589} (\bibinfo {year}
  {2023}{\natexlab{c}})},\ \Eprint {https://arxiv.org/abs/2306.02324}
  {arXiv:2306.02324 [gr-qc]} \BibitemShut {NoStop}%
\bibitem [{\citenamefont {Wu}(2023{\natexlab{d}})}]{Wu:2023meo}%
  \BibitemOpen
  \bibfield  {author} {\bibinfo {author} {\bibfnamefont {D.}~\bibnamefont
  {Wu}},\ }\bibfield  {title} {\bibinfo {title} {{Topological classes of
  thermodynamics of the four-dimensional static accelerating black holes}},\
  }\href {https://doi.org/10.1103/PhysRevD.108.084041} {\bibfield  {journal}
  {\bibinfo  {journal} {Phys. Rev. D}\ }\textbf {\bibinfo {volume} {108}},\
  \bibinfo {pages} {084041} (\bibinfo {year} {2023}{\natexlab{d}})},\ \Eprint
  {https://arxiv.org/abs/2307.02030} {arXiv:2307.02030 [hep-th]} \BibitemShut
  {NoStop}%
\bibitem [{\citenamefont {Wu}\ \emph {et~al.}(2024{\natexlab{b}})\citenamefont
  {Wu}, \citenamefont {Gu}, \citenamefont {Zhu}, \citenamefont {Jiang},\ and\
  \citenamefont {Yang}}]{Wu:2024rmv}%
  \BibitemOpen
  \bibfield  {author} {\bibinfo {author} {\bibfnamefont {D.}~\bibnamefont
  {Wu}}, \bibinfo {author} {\bibfnamefont {S.-Y.}\ \bibnamefont {Gu}}, \bibinfo
  {author} {\bibfnamefont {X.-D.}\ \bibnamefont {Zhu}}, \bibinfo {author}
  {\bibfnamefont {Q.-Q.}\ \bibnamefont {Jiang}},\ and\ \bibinfo {author}
  {\bibfnamefont {S.-Z.}\ \bibnamefont {Yang}},\ }\bibfield  {title} {\bibinfo
  {title} {{Topological classes of thermodynamics of the static multi-charge
  AdS black holes in gauged supergravities: novel temperature-dependent
  thermodynamic topological phase transition}},\ }\href
  {https://doi.org/10.1007/JHEP06(2024)213} {\bibfield  {journal} {\bibinfo
  {journal} {JHEP}\ }\textbf {\bibinfo {volume} {06}},\ \bibinfo {pages}
  {213}},\ \Eprint {https://arxiv.org/abs/2402.00106} {arXiv:2402.00106
  [hep-th]} \BibitemShut {NoStop}%
\bibitem [{\citenamefont {Wu}\ \emph {et~al.}(2024{\natexlab{c}})\citenamefont
  {Wu}, \citenamefont {Liu}, \citenamefont {Wu},\ and\ \citenamefont
  {Mann}}]{Wu:2024asq}%
  \BibitemOpen
  \bibfield  {author} {\bibinfo {author} {\bibfnamefont {D.}~\bibnamefont
  {Wu}}, \bibinfo {author} {\bibfnamefont {W.}~\bibnamefont {Liu}}, \bibinfo
  {author} {\bibfnamefont {S.-Q.}\ \bibnamefont {Wu}},\ and\ \bibinfo {author}
  {\bibfnamefont {R.~B.}\ \bibnamefont {Mann}},\ }\bibfield  {title} {\bibinfo
  {title} {{Novel Topological Classes in Black Hole Thermodynamics}},\
  }\href@noop {} {\  (\bibinfo {year} {2024}{\natexlab{c}})},\ \Eprint
  {https://arxiv.org/abs/2411.10102} {arXiv:2411.10102 [hep-th]} \BibitemShut
  {NoStop}%
\end{thebibliography}
%

\end{document}